\documentclass[prb,aps,showpacs]{revtex4}
\usepackage{amssymb}
\usepackage{amsmath}

\advance\textheight by 0.14in
\advance\topmargin by -0.1in

\def\q{{\bf q}}

\def\r{{\bf r}}

\def\K{{\bf K}}

\def\ltsim{\vbox {\hbox{\lower .8\baselineskip \hbox{$<$}} \break
                 \hbox{\lower 0.2\baselineskip \hbox{$\sim$}} } }
\def\gtsim{\vbox {\hbox{\lower .8\baselineskip \hbox{$>$}} \break
                 \hbox{\lower 0.2\baselineskip \hbox{$\sim$}} } }

\begin{document}

\title{Dynamical Conductivity of Disordered Quantum Hall Stripes}
\author{Mei-Rong Li$^{1}$, H.A. Fertig$^{2,3}$, R. C\^{o}t\'{e}$^{1}$, 
Hangmo Yi$^{4}$}
\affiliation{$^{1}$D\'epartement de Physique, Universit\'e de Sherbrooke, 
Sherbrooke, Qu\'ebec, Canada J1K 2R1\\
$^{2}$Department of Physics and Astronomy, University of Kentucky,
Lexington, Kentucky 40506-0055\\
$^{3}$Department of Physics, Indiana University, Bloomington, IN 47405\\
$^{4}$Department of Physics, Soongsil University, Seoul, 156-743, Korea}
\date{\today}

\begin{abstract}
We present a detailed theory for finite-frequency conductivities Re$%
[\sigma_{\alpha\beta}(\omega)]$ of quantum Hall stripes, which form at Landau
level $N\geq 2$ close to half filling, in the presence of weak Gaussian
disorder. We use an effective elastic theory to describe the low-energy
dynamics of the stripes with the dynamical matrix being determined through
matching the density-density correlation function obtained in the
microscopic time-dependent Hartree-Fock approximation. We then apply
replicas and the Gaussian variational method to deal with the disorder.
Within this method, a set of saddle point equations for the retarded self
energies are obtained, which are solved numerically to get Re$%
[\sigma_{\alpha\beta}(\omega)]$. We find a quantum depinning transition as $%
\Delta\nu$, the fractional part of the filling factor, approaches a critical
value $\Delta\nu_c$ from below. For $\Delta\nu<\Delta\nu_c$, the pinned
state is realized in a replica symmetry breaking (RSB) solution, and the
frequency-dependent conductivities in both the directions perpendicular and
parallel to the stripes show resonant peaks. These peaks shift to zero
frequency as $\Delta\nu\rightarrow \Delta\nu_c$. For $\Delta\nu\ge\Delta%
\nu_c $, we find a \emph{partial RSB (PRSB)} solution in which there is RSB
perpendicular to the stripes, but replica symmetry along the stripes,
leading to free sliding along the stripe direction. The quantum depinning
transition is in the Kosterlitz-Thouless universality class. The result is
consistent with a previous renormalization group analysis.
\end{abstract}

\maketitle


\section{Introduction}

Charge density waves (CDWs) may form in many correlated electronic systems
when the Coulomb interaction dominates over the kinetic energy. For a 
two-dimensional electron gas in a perpendicular magnetic field, the quantization
of the kinetic energy into Landau levels can enhance this possibility \cite%
{stripestheory}. Each Landau level is highly degenerate, with the number of
states equal to the number of magnetic flux quanta passing through the
system. If the field and corresponding degeneracy is sufficiently large, the
low-energy physics of the system may then be dominated by electrons in the
highest partially occupied ($N$th) Landau level, with the other electrons
essentially renormalizing the effective Coulomb interaction in this level 
\cite{AG}. In this situation, the kinetic energy is quenched and the system
arranges itself in order to minimize the interactions. The Hartree-Fock
approximation \cite{stripestheory} predicts the formation of CDW ground
states for $N\geq 2$. These CDWs evolve from Wigner crystals to
\textquotedblleft bubble states\textquotedblright\ \cite{stripestheory} as
the partial Landau-level filling factor $\Delta \nu $ increases. For $\Delta
\nu \gtrsim 0.4$, the bubble states give way to stripe states (also
called unidirectional CDWs.) This Hartree-Fock result is corroborated by
density matrix renormalization group calculations \cite{DMRG} and exact
diagonalization studies \cite{exactdiag}. DC transport experiments indeed
observe highly anisotropic, and apparently metallic, conductivity \cite%
{stripesexpr} near half-filling of higher Landau levels ($N\geq 2$). This is
likely due to the formation of stripe states.

Because the stripe state breaks translational symmetry in only one
direction, it has the symmetry of a smectic liquid crystal \cite{fradkin},
and as in that system supports a set of gapless phonon modes \cite{CF}.
These modes are present because the stripe state lacks any restoring force
when a single stripe slides with respect to the others. 
In the context of an electron system, the smectic state can
be thought of as a self-organized array of Luttinger liquids \cite%
{fradkin,edgestatemodel}, a state of fermions that does not obey Landau
Fermi liquid theory and (so far) is known to exist only in one dimension 
\cite{voit}. One way of viewing the Luttinger liquid is in terms of a one
dimensional crystal that has lost long-range order due to quantum
fluctuations \cite{kolomeisky}. This idea is easily generalized to the case
of stripes \cite{fertig99}, and suggests that the low-energy degrees of
freedom for the system may be described in terms of a displacement field.
This will be the basic language for our study. 

It has long been recognized that disorder can pin a CDW and render it
insulating \cite{FL1,Larkin}. (Similarly, Luttinger liquids may be pinned by
impurities in spite of their liquid-like correlations \cite{KF}.) In this
situation, the real part of the zero wavevector, finite-frequency
conductivity $\sigma _{\alpha \beta }(\omega )$ vanishes as $\omega
\rightarrow 0$, and has a resonance at higher $\omega $ with a peak (or
pinning) frequency and width that are determined by the effective restoring
force due to the disorder \cite{FL1}. Such behavior has indeed been observed
in high Landau levels for $\Delta \nu $ sufficiently far away from 1/2 so
that one expects the electrons to be organized into bubble states\cite%
{muwave}. As $\Delta \nu $ is increased, there is a general trend for the
pinning frequency to decrease, eventually becoming lost in the noise as the
filling factor approaches the value at which the DC conductivity becomes
anisotropic and metallic \cite{stripesexpr}. Experiments to better resolve
the dynamical conductivity as the stripe phase is entered are currently
underway \cite{Florida}.

A fundamental question that arises in this context is whether the apparent
metallic behavior seen in DC transport experiments represents the true zero
temperature behavior in the stripe state. While current data suggests the
diagonal conductivities $\sigma _{xx}$ and $\sigma _{yy}$ saturate to finite
values at low temperatures, presumably such experiments can answer this
fundamental question
unambiguously only by reaching significantly lower temperatures than are
currently available. The possibility that near half-filling the stripes may
not be fully pinned is extremely intriguing because, if this is indeed the
case, then the electrons have avoided becoming localized and the resulting
anisotropic metal cannot be a Fermi liquid. Thus this state could well be a
higher dimensional analogue of a Luttinger liquid. Developing and understanding 
the results of experiments beyond DC transport -- such as measurements of the 
dynamical conductivity \cite{muwave} -- then take on an additional significance.

One possible route to metallic behavior for the stripe system could be via a
depinning transition. In principle this could happen in a one dimensional
Luttinger liquid, if one could continuously tune the interactions from
repulsive to attractive \cite{KF}. Because the stripe system has a larger
parameter space needed to describe its elastic properties than the single
stiffness that describes a one-dimensional solid (specifically, one needs to
estimate the dynamical matrix along a line in the Brillouin zone, as we
discuss below), it is possible to arrive at this depinning transition even
when the bare interaction parameters among the electrons are purely
repulsive \cite{fertig99}. The question of whether stripes may become
depinned must be answered via a detailed calculation of the stripe
elasticity, and depending on how one estimates this, different answers are
possible \cite{YFC,edgestatemodel}, as we discuss in more detail below.

In what follows, we will adopt an approach that models the quantum Hall (QH)
stripe system as an array of one dimensional, quantum disordered solids as
shown in Fig.~\ref{stripeselastic}. We
estimate the dynamical matrix of this system by matching the elastic theory
to the results of a microscopic calculation within the time-dependent
Hartree-Fock approximation (TDHFA). This approach was taken by some of us 
\cite{YFC} in a perturbative renormalization group (RG) calculation, which
demonstrated that a quantum depinning transition can occur as $\Delta\nu
\rightarrow \Delta\nu_c$ from below, with the critical partial filling $%
\Delta\nu_c$ depending on the details of the system: Landau level index,
layer thickness, disorder strength, etc. Our goal is to compute the
dynamical conductivity as the system passes through the transition, to
identify signatures that would indicate that the system has passed into an
unpinned stripe state. A brief summary of our major results has been
published elsewhere \cite{LFCY}. In this article, we provide details of the
calculations as well as some further results.

Our general approach to this problem is to use replicas \cite{MPV} and the
Gaussian variational method (GVM), as was first introduced by M\'{e}zard and
Parisi for elastic manifolds \cite{MP} and then further developed by
Giamarchi and Le Doussal and their coworkers in applications to a variety of
condensed matter systems \cite{GLD96,GLD95,GO,CGLD}. In the QH systems, 
this method was used by Chitra {\it et al.} \cite{CGLD} for 
pinned Wigner crystals in the $N=0$ Landau level, and by Orignac and Chitra 
\cite{OC} for stripes, in the latter case using a different set of approximations 
than us and yielding very different results than those described below.
Because of the strong fluctuations inherent in the depinning transition, we 
have found that one cannot correctly solve for the dynamical conductivity using 
the \textquotedblleft semiclassical approximation\textquotedblright\ to the
saddle point equations (SPE's), which we review below, that has given this
approach its attractive simplicity. By relaxing this approximation we will
see that in the depinned state the dynamical conductivity can have a
surprising power-law frequency dependence, and a discontinuous behavior at
the transition that is analogous to the universal stiffness jump that occurs
at a Kosterlitz-Thouless (KT) transition \cite{nelson}. Moreover, we will see
that the solution to the SPE's that yield this behavior have an unusual
structure involving breaking the replica symmetry for motion perpendicular
to the stripes, while preserving it parallel to the stripes. This \textit{%
partial replica symmetry breaking} (PRSB) indicates that the stripes may be
pinned for perpendicular motion while free to slide relative to one another.
This qualitative behavior was anticipated by the perturbative RG study \cite%
{YFC}.

As discussed above, when the system is pinned there are resonant peaks 
which appear in $\sigma _{xx}$ and $\sigma
_{yy}$. (We choose the $\hat{x}$ direction to be perpendicular to the
stripes, and the $\hat{y}$ direction to be parallel to it as shown in 
Fig.~\ref{stripeselastic}.
Of course, the two diagonal conductivities are not the same due to the
anisotropy of the stripe state.) The peaks drop to zero frequency as $%
\Delta \nu _{c}$ is approached from below, with their weights
increasing for motion along the stripes, and decreasing for motion
perpendicular to them. As the transition is approached, the resonance peaks
become increasingly asymmetric. Upon crossing the transition, 
$\sigma _{yy}$ develops a $\delta $-function at $\omega =0$, indicating
superconducting behavior, while $\sigma _{xx}$ rises from zero as a power of 
$\omega $. This unusual behavior is a result of the power-law correlations
associated with the Luttinger liquid-like behavior of the unpinned state. We
note that our $\omega =0$ results are not consistent with the DC
conductivity results seen in experiments, although preliminary experimental
results for finite frequency do bear some resemblance to our predictions 
\cite{Florida}. We will comment below on what is missing from our model that
we believe leads to this discrepancy.

This paper is organized as follows. In Sec.~II we review the procedure for
determining the dynamical matrix in the elastic model. This is followed by a
review in Sec.~III of the qualitative effects of disorder within the RG
analysis. In Sec.~IV, we review the replica and GVM which leads to a set of
saddle point equations (SPE's) for the self energy. Solutions of the SPE's
and the result for the conductivities are presented in Sec.~V, focusing on
the pinned state for $\Delta \nu <\Delta \nu _{c}$, and in Sec.~VI, which is
devoted to the depinned state for $\Delta \nu >\Delta \nu _{c}$. We discuss
the nature of the depinning transition in Sec.~VII, and conclude in
Sec.~VIII. There are four appendices: the first summarizes the Hartree-Fock 
(HF) and the TDHFA formalisms, the result of which is used for determination 
of the dynamical matrix; the second gives a derivation for the
inversion rules needed for hierarchical matrices of the type dealt with in
this paper; the third discusses analytic continuation of the dynamical
conductivity from imaginary time to real time; and the forth discusses
another possible solution to the SPE's that has unphysical properties.

\section{Elastic model of QH stripes}

\subsection{Elastic action}

\begin{figure}[h]
\begin{picture}(200,140)
\leavevmode\centering\includegraphics{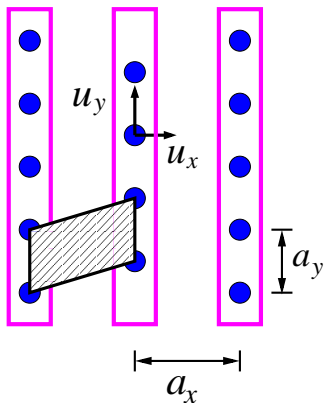}
\end{picture}
\caption{Schematic diagram of quasi one-dimensional arrays modelling the
QH stripes. The stripes are along the $\hat{y}$ direction, and $u_x$ and 
$u_y$ are the displacement fields. The shaded area is the unit cell
of the stripes crystal with the volume $a_xa_y$.}
\label{stripeselastic}
\end{figure}

In our approach, low energy distortions from the mean-field state are
described by an elastic model, with displacement fields $u_{x}\left( \mathbf{%
r}\right) $ and $u_{y}\left( \mathbf{r}\right) $ representing the effective
dynamical variables of the QH stripes. Fig.~\ref{stripeselastic} shows
schematically the one-dimensional arrays modelling the stripes. 
They obey single Landau level
dynamics \cite{kubo} $\left[ u_{x}(\mathbf{R}),u_{y}(\mathbf{R}^{\prime })%
\right] =il_{B}^{2}\delta_{\mathbf{R},\mathbf{R}^{\prime }} $,
where $l_{B}=\sqrt{\hbar c/eB}$ is the magnetic length. In the pure limit,
the Euclidean action of the elastic model may be written as (throughout this
work, we use the unit $k_{B}=\hbar =1$) 
\begin{equation}
S_{0}={\frac{1}{2T}}\sum_{\mathbf{q},\omega _{n}}\sum_{\alpha ,\beta
=x,y}u_{\alpha }\left( \mathbf{q},\omega _{n}\right) \,G_{\alpha \beta
}^{(0)-1}\left( \mathbf{q},\omega _{n}\,\right) u_{\beta }\left( -\mathbf{q}%
,-\omega _{n}\right) ,  \label{pureS0}
\end{equation}%
where $T$ is the temperature, $\omega _{n}\left( =2\pi n/T\right) $ the
bosonic Matsubara frequency, and 
\begin{equation}
G^{(0)}_{\alpha \beta}\left( \mathbf{q},i\omega _{n}\right) 
=\frac{l_{B}^{4}}{\left( \omega_{n}^{2}+\omega _{\mathbf{q}}^{2}\right) }\left( 
\begin{array}{cc}
D_{yy}\left( \mathbf{q}\right) & \frac{\omega _{n}}{l_{B}^{2}}-D_{xy}\left( 
\mathbf{q}\right) \\ 
-\frac{\omega _{n}}{l_{B}^{2}}-D_{yx}\left( \mathbf{q}\right) & D_{xx}\left( 
\mathbf{q}\right)%
\end{array}%
\right)_{\alpha \beta}  \label{GF0}
\end{equation}%
is the unperturbed Green's function of the displacement fields with 
$D_{\alpha\beta}(\q)$ being the dynamical matrix and 
\begin{equation}
\omega _{\mathbf{q}}=l_{B}^{2}\sqrt{D_{xx}\left( \mathbf{q}\right)
D_{yy}\left( \mathbf{q}\right) -D_{xy}^{2}\left( \mathbf{q}\right) }
\label{eigenmode}
\end{equation}%
being a general expression for phonon modes of a charged elastic system
in a strong magnetic field (magnetophonon modes).
As always for a Gaussian theory, the correlation function may be
expressed in terms of the Green's function via 
\begin{equation}
G_{\alpha \beta }^{(0)}\left( \mathbf{q},\omega _{n}\right)
=\int_{0}^{1/T}d\tau e^{i\omega _{n}\tau }\,\left\langle \mathcal{T}_{\tau
}u_{\alpha }\left( \mathbf{q},\tau \right) u_{\beta }\left( -\mathbf{%
q},0\right) \right\rangle _{S_{0}},
\end{equation}%
where $\left\langle \cdots \right\rangle _{S_{0}}$ denotes an average over
the displacement fields with the usual weighting factor $e^{-S_{0}}$, and 
$\mathcal{T}_{\tau }$ is the imaginary time ordering operator.

Because of inversion and reflection symmetries, and the fact that the
dynamical matrix elements in real space are real, we have  
\begin{eqnarray}
&&D_{\alpha \beta }\left( \mathbf{q}\right) =D_{\beta \alpha }\left( -%
\mathbf{q}\right) , \\
&&D_{xy}\left( \mathbf{q}\right) =D_{yx}\left( \mathbf{q}\right) , \\
&&D_{xy}\left( q_{x},q_{y}\right) =-D_{xy}\left( -q_{x},q_{y}\right)
=-D_{xy}\left( q_{x},-q_{y}\right) ,
\end{eqnarray}%
so that the unperturbed Green's function has the symmetries 
\begin{eqnarray}
&&G_{\alpha \alpha }^{(0)}\left( \mathbf{q},\omega _{n}\right) =G_{\alpha
\alpha }^{(0)}\left( -\mathbf{q},\omega _{n}\right) =G_{\alpha \alpha
}^{(0)}\left( \mathbf{q},-\omega _{n}\right) =G_{\alpha \alpha }^{(0)}\left(
-\mathbf{q},-\omega_{n}\right) ,  \label{symmetry1} \\
&&G_{xy}^{(0)}\left( \mathbf{q},\omega _{n}\right) =G_{yx}^{(0)}\left( 
\mathbf{q},-\omega_{n} \right),  \label{symmetry2} \\
&&G_{xy}^{(0)}\left( q_{x},q_{y},\omega _{n}\right) =G_{xy}^{(0)}\left(
-q_{x},-q_{y},\omega _{n}\right) =-G_{xy}^{(0)}\left( -q_{x},q_{y},\omega
_{n}\right) =-G_{xy}^{(0)}\left( q_{x},-q_{y},\omega _{n}\right) .
\label{symmetry3}
\end{eqnarray}

To perform quantitative calculations, it is necessary to produce estimates
of the dynamical matrix elements $D_{\alpha \beta }(\mathbf{q})$ for the
QH stripe states. We do this with a matching procedure that uses
results from microscopic TDHFA computations. Below we briefly review this
matching procedure.

\subsection{Relation between $G^{(0)}_{\alpha\beta}
\left( \mathbf{q},\protect\omega \right) $
and guiding-center density-density correlation function}

In a classical model, each site of the crystal is occupied by an electron
whose charge density is specified by a form factor $f\left( \mathbf{r}%
\right) $ (with $\int d\mathbf{r}f\left( \mathbf{r}\right) =1\,$). In the
absence of any fluctuations these electrons will lie on the oblique Bravais
lattice as shown in Fig.~\ref{stripeselastic}.
Fluctuations around this reference state are
given in terms of the displacement fields $\mathbf{u}(\mathbf{R})$. The
time-dependent electronic density is then written as
\begin{equation}
n\left( \mathbf{r},t\right) =\sum_{\mathbf{R}}f\left( \mathbf{r}-\mathbf{R}-%
\mathbf{u}\left( \mathbf{R},t\right) \right) .  \label{21_5}
\end{equation}%
The Fourier transform of this density is given by%
\begin{equation}
n(\mathbf{q},t)=\int d\mathbf{r}e^{-i\mathbf{q}\cdot \mathbf{r}}n\left( 
\mathbf{r},t\right) \approx f(\mathbf{q})\delta _{\mathbf{q},\mathbf{K}%
}-if(\mathbf{q})\sqrt{N_{s}}\mathbf{q}\cdot \mathbf{u}(\mathbf{q}),
\label{1_17}
\end{equation}%
where $N_{s}$ is the number of crystal sites or electrons and 
$\K$ is a reciprocal lattice vector. The form factor $%
f(\mathbf{r})$ is real and has inversion symmetry so that $f(\mathbf{q})$ is
real.

The fact that the density fluctuations are related to the displacement field 
via %
\begin{equation}
\delta n(\mathbf{q}+\mathbf{K},t)\approx -if(\mathbf{q+K})\sqrt{N_{s}}\left( 
\mathbf{q+K}\right) \cdot \mathbf{u}(\mathbf{q})
\end{equation}%
(with $\q$ a vector in the first Brillouin zone of the reciprocal lattice) 
implies that we can relate the displacement Green's function
$G^{(0)}_{\alpha\beta}\left( \mathbf{q},\protect\omega \right) $
to the density-density correlation function 
$\chi _{\mathbf{K},\mathbf{K}^{\prime }}^{\left( n,n\right) }\left( \mathbf{q}%
,\tau \right)$ (introduced in Appendix~\ref{HFappendix}) through
\begin{equation}
\chi _{\mathbf{K},\mathbf{K}^{\prime }}^{\left( n,n\right) }\left( \mathbf{q}%
,\tau \right) =-N_{s}f(\mathbf{q+K})f\left( \mathbf{q}+\mathbf{K}^{\prime
}\right) \left[ \left( \mathbf{q+K}\right) \cdot \widehat{G}^{(0)}\left( \mathbf{q}%
,\tau \right) \cdot \left( \mathbf{q+K}^{\prime }\right) \right] .
\label{chiGF}
\end{equation}%
Here $\chi _{\mathbf{K},\mathbf{K}^{\prime }}^{\left( n,n\right) }
\left( \mathbf{q},\tau \right)$ is a quantity that we compute in the 
microscopic TDHFA \cite{CF}. In Appendix~\ref{HFappendix}, we summarize the HF and TDHF 
formalisms. Eq.~(\ref{21_13}) there will be used for the determination of 
the dynamical matrix. Substituting Eq.~(\ref{GF0}) in Eq.~(\ref{chiGF}) yields 
\begin{equation}
\chi _{\mathbf{K},\mathbf{K}^{\prime }}^{\left( n,n\right) }\left( \mathbf{q}%
,i\omega _{n}\right) =-\frac{N_{s}l_{B}^{4}}{\left( \omega _{n}^{2}+\omega _{%
\mathbf{q}}^{2}\right) }\left[ \Gamma _{1}+\Gamma _{2}\frac{\omega _{n}}{%
l_{B}^{2}}\right] f\left( \mathbf{q}+\mathbf{K}\right) f\left( \mathbf{q}+%
\mathbf{K}^{\prime }\right) , \label{chiwn} 
\end{equation}%
with the definitions 
\begin{equation}
\Gamma _{1}=-\left( \mathbf{q}+\mathbf{K}\right) \times \overleftrightarrow{D%
}\left( \mathbf{q}\right) \times \left( \mathbf{q}+\mathbf{K}^{\prime
}\right) ,
\end{equation}%
and 
\begin{equation}
\Gamma _{2}=\left( \mathbf{q}+\mathbf{K}\right) \times \left( \mathbf{q}+%
\mathbf{K}^{\prime }\right) .
\end{equation}%
The two-dimensional vector product in the last two equations stands for $%
\mathbf{a}\times \mathbf{b}=a_{x}b_{y}-a_{y}b_{x}.$

The analytical continuation of $\chi _{\mathbf{K},\mathbf{K}^{\prime
}}^{\left( n,n\right) }\left( \mathbf{q},i\omega _{n}\right) $ in 
Eq.~(\ref{chiwn}) results in 
\begin{equation}
\chi _{\mathbf{K},\mathbf{K}^{\prime }}^{\left( n,n\right) }\left( \mathbf{q}%
,\omega \right) =-N_{s}l_{B}^{4}\left[ \frac{Z}{\omega +i\delta +\omega _{%
\mathbf{q}}}-\frac{Z^{\ast }}{\omega +i\delta -\omega _{\mathbf{q}}}\right]
f\left( \mathbf{q}+\mathbf{K}\right) f\left( \mathbf{q}+\mathbf{K}^{\prime
}\right) ,  \label{21_8}
\end{equation}%
where 
\begin{equation}
Z=\frac{\Gamma _{1}}{2\omega _{\mathbf{k}}}-i\frac{\Gamma _{2}}{2l_{B}^{2}}.
\end{equation}%
\qquad

We can now request that Eq.~(\ref{21_8}) be equivalent to  
Eq.~(\ref{21_13}) in Appendix~\ref{HFappendix} in order to obtain the 
dynamical matrix. This requires that 
\begin{equation}
\Delta \nu \, l_{B}^{4}Z^{\ast }f\left( \mathbf{q}+\mathbf{K}\right) f\left( 
\mathbf{q}+\mathbf{K}^{\prime }\right) =F\left( \mathbf{q}+\mathbf{K}\right)
F\left( \mathbf{q}+\mathbf{K}^{\prime }\right) W_{i}\left( \mathbf{q}+%
\mathbf{K},\mathbf{q}+\mathbf{K}^{\prime }\right) ,  \label{21_10}
\end{equation}%
where $W_{i}\left( \mathbf{q}+\mathbf{K},\mathbf{q}+\mathbf{K}^{\prime
}\right) $ is the weight associated with the magnetophonon frequency $%
\varepsilon _{i}$ in the TDHFA\ response and $\Delta \nu =N/N_{\varphi }$ is
the filling factor of the partially filled level. The magnetophonon
frequency $\varepsilon _{i}$ is found, at small wavevector $\mathbf{q}$, by
locating the eigenvalue $\varepsilon _{i}$ of the matrix $M$ defined in Eq. (%
\ref{eqnm}) with the biggest weight $W_{i}\left( \mathbf{q}+\mathbf{K},%
\mathbf{q}+\mathbf{K}\right) $ in the {\it diagonal} response function $%
\chi _{\mathbf{K},\mathbf{K}}^{\left( n,n\right) }\left( \mathbf{q},\omega
\right) .$

A careful examination shows that, because $\omega _{\mathbf{q}}$ is given by
the determinant of the matrix $\hat{D}$, the quantity $\Gamma _{1}/2\omega _{%
\mathbf{q}}$ is unchanged if all the components of the dynamical matrix are
multiplied by some constant. Eq.~(\ref{21_10}) is thus indeterminate. To
avoid this, we replace $\omega _{\mathbf{q}}$ by $\varepsilon _{i}$ in this
equation. Our final equation is then%
\begin{equation}
f\left( \mathbf{q}+\mathbf{K}\right) f\left( \mathbf{q}+\mathbf{K}^{\prime
}\right) \left[ l_{B}^{2} \Gamma _{1}+i\varepsilon _{i}\Gamma _{2}\right] 
=\frac{%
2\varepsilon _{i}}{\Delta \nu l_{B}^{2}}F\left( \mathbf{q}+\mathbf{K}\right) 
F\left( \mathbf{%
q}+\mathbf{K}^{\prime }\right) W_{i}\left( \mathbf{q}+\mathbf{K},\mathbf{q}+%
\mathbf{K}^{\prime }\right) .  \label{1_14}
\end{equation}%
With this equation, we can determine the 3 components of the dynamical
matrix as well as the form factors $f\left( \mathbf{q}+\mathbf{K}\right) .$

\subsection{Matching procedure}

At this point, it is worthwhile remarking that, in the HFA, there is an
extremely small energy difference (of order $10^{-6}$ $e^{2}/\kappa l_{B}$)
between the energies of the stripe crystals with in-phase and out-of-phase
modulation on adjacent stripes. As a result, the magnetophonon dispersion in
the TDHFA\ has a very small gap in the perpendicular direction. This
interstripe locking energy is, however, not accessible within our numerical
accuracy so that our calculated magnetophonon dispersion is that appropriate
for a smectic. In particular, it contains a line of gapless modes for $%
q_{x}\neq 0,q_{y}=0$. Because of this nodal line, we need to fit the
dynamical matrix for small $q_{y}$ i.e., for long wavelengths along the
stripes, and for all values of $q_{x}$ in the Brillouin zone. Indeed, these
low-energy modes play a crucial role in determining the effects of both
quantum and thermal fluctuations on the system.

We choose to solve Eq. (\ref{1_14}) for the shortest three reciprocal
lattice vectors: $\mathbf{K,K}%
^{\prime }=(0,0),(0,\pm K_{y0})$ where $K_{y0}=2\pi /a_{y}$ with 
$a_y$ being the lattice constant along the stripes direction (see 
Fig.~\ref{stripeselastic}). For each $%
\mathbf{q}$, the TDHFA calculation provides ten independent numbers, nine in
the $3\times 3$ Hermitian matrix $W_{i}\left( \mathbf{q}+\mathbf{K},\mathbf{q%
}+\mathbf{K}^{\prime }\right) $ and one in $\varepsilon _{i}$. We use six of
them to determine $D_{xx}$, $D_{xy}$, $D_{yy}$, and the three real
parameters $f(\mathbf{q}+\mathbf{K})$. The rest may be used to check the
consistency of the numerical procedure. The final result \cite{YFC} indicates
that the matching is very accurate.

\begin{figure}[h]
\begin{picture}(250,200)
\leavevmode\centering\includegraphics{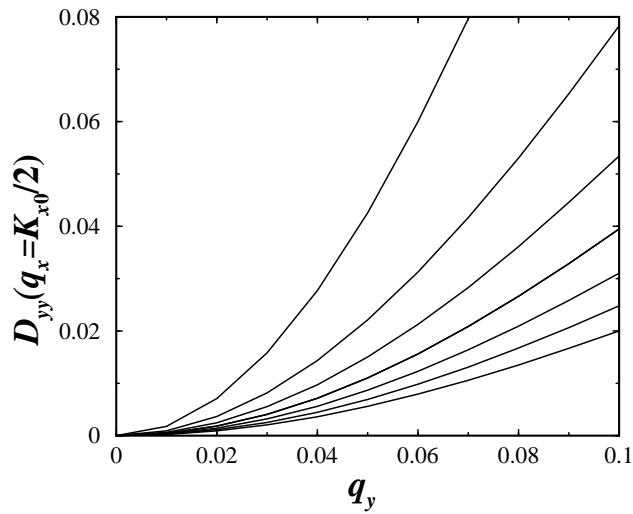}
\end{picture}
\caption{Dynamical matrix $D_{yy}$ as a function of $q_{y}$ at small $q_{y}$
and $q_{x}=K_{x0}/2$. Curves from right to left correspond to $\Delta 
\protect\nu =0.36,0.38,0.4,0.42,0.44,0.46,0.48$, respectively. }
\label{Dyy}
\end{figure}

A typical result for $D_{yy}(\mathbf{q})$ as a function of $q_{y}$ at small $%
q_{y}$ and $q_{x}=K_{x0}/2$ is shown in Fig.~\ref{Dyy}. Clearly $D_{yy}(%
\mathbf{q})$ is quadratic in $q_{y}$. Indeed, based on symmetry
considerations \cite{edgestatemodel,CF}, the low energy sector of $\mathbf{D}%
(\mathbf{q})$ should have the form for small $q_{y}$: 
\begin{eqnarray}
&&D_{xx}\left( \mathbf{q}\right) \simeq d_{xx}\left( q_{x}\right) +\kappa
_{b}q_{y}^{4},  \label{smecticDxx} \\
&&D_{xy}\left( \mathbf{q}\right) \simeq d_{xy}\left( q_{x}\right) q_{y},
\label{smecticDxy} \\
&&D_{yy}\left( \mathbf{q}\right) \simeq d_{yy}\left( q_{x}\right) q_{y}^{2},
\label{smecticDyy}
\end{eqnarray}%
where $\kappa _{b}$ is the bending coefficient. The absence of a quadratic $%
q_{y}$ term in $D_{xx}$ follows from rotational symmetry and is the major
difference between a smectic and a crystal dynamical matrix. 
In our calculation below, we will use this smectic form,
determining $d_{xx}\left( q_{x}\right) $, $d_{xy}\left( q_{x}\right) $ and $%
d_{yy}\left( q_{x}\right) $ on a grid of $q_{x}$ points numerically.
The $\kappa_{b}q_{y}^{4}$ in Eq.~(\ref{smecticDxx}) reflects the bending energy 
of the stripes. In practice, this term merely plays the role of high-$q_y$ cutoff 
and thus $\kappa_{b}$ is chosen for convenience to be 2 in our numerical
calculation. 

Inserting Eqs.~(\ref{smecticDxx}-\ref{smecticDyy}) into Eq.~(\ref{eigenmode}%
) yields 
\begin{equation}
\omega _{\mathbf{q}}\simeq l_{B}^{2}\sqrt{d_{xx}\left( q_{x}\right)
d_{yy}\left( q_{x}\right) -d_{xy}^{2}\left( q_{x}\right) }\;q_{y}
\end{equation}
for small $q_y$.

Once the Green's function has been determined, we can easily compute the
conductivity. Since the electric current is carried by the charge, the
current density can be expressed as 
\begin{equation}
\mathbf{j}\left( \mathbf{q},\tau \right) =ie
{\frac{d\mathbf{u}\left( \mathbf{q},\tau \right) }{d\tau }}.
\end{equation}%
The conductivity is then determined by the Kubo formula to be 
\begin{equation}
\sigma _{\alpha \beta }\left( \omega \right) =-{\frac{1}{\omega \,a_x a_y}}\,
\left[\int_{0}^{1/T}d\tau e^{i\omega _{n}\tau }\left\langle j_{\alpha }\left( 
\mathbf{q}=0,\tau \right) j_{\beta }\left( \mathbf{q}=0,0\right)
\right\rangle \right] _{i\omega _{n}\rightarrow \omega +i0^{+}}=-{\frac{e^{2}%
}{a_{x}a_{y}}}\,i\omega \,G_{\alpha \beta }^{\mathrm{ret}}\left( \mathbf{q}%
=0,\omega \right) .  \label{conductivity0}
\end{equation}%
where $a_x$ is the distance between the centers of two neighboring stripes
(see Fig.~\ref{stripeselastic}).
It is easy to check that in the pure limit, the electromagnetic response 
of the system is purely transverse. Calculating $G_{\alpha \beta}^{\mathrm{ret}}$ 
in the presence of disorder is our next (and indeed most important) task.

\section{Qualitative effect of disorder}

\subsection{Modeling the disorder}

We assume that the disorder can be modeled as a Gaussian random potential $V(%
\mathbf{r})$. The disorder action reads 
\begin{equation}
S_{\mathrm{imp}}=\int d\mathbf{r}\,\int_{0}^{1/T}d\tau \,V\left( \mathbf{r}%
\,\right) n\left( \mathbf{r},\tau \right)  \label{1_15}
\end{equation}%
where $V\left( \mathbf{r}\right) $ has the following Gaussian distribution
function 
\begin{equation}
P\left( V\right) =\exp \left[ -{\frac{1}{2}} \int d\r_1 \int d\r_2
V\left( \mathbf{r}_{1}\right) \Gamma ^{-1}\left( \mathbf{r}_{1}-\mathbf{%
r}_{2}\right) V\left( \mathbf{r}_{2}\right) \right] ,
\end{equation}%
with 
\begin{equation}
\Gamma \left( \mathbf{r}_{1}-\mathbf{r}_{2}\right) =\overline{V\left( 
\mathbf{r}_{1}\right) V\left( \mathbf{r}_{2}\right) }=V_{0}^{2}\,a_{x}%
\,a_{y}\,\delta \left( \mathbf{r}-\mathbf{r}^{\prime }\right) .
\end{equation}%
Here the overline denotes average over disorder: 
\begin{equation}
\overline{A}={\frac{\int \mathcal{D}VP\left( V\right) A}{\int \mathcal{D}%
VP\left( V\right) }}.
\end{equation}%
The electron density operator $n\left( \mathbf{r},\tau \right) $ in 
Eq.~({\ref{1_15}}) must, in order to capture the possibility of pinning by disorder,
be approximated more accurately than was needed in the matching procedure discussed
in the preceding section.
Following Giamarchi and Le Doussal \cite{GLD95}, under the assumption of
small $\nabla {\mathbf{u}}(\mathbf{r})$ (which is justified for weak
disorder) we write 
\begin{equation}
n(\mathbf{q},\tau )\simeq f(\mathbf{q})\left[ N_s-i\sqrt{N_s}
\mathbf{q}\cdot \mathbf{u}%
\left( \mathbf{q},\tau \right) +\sum_{\mathbf{K}\neq 0}\int d\mathbf{r}\,e^{i%
\mathbf{K}\cdot \left[ \mathbf{r}-\mathbf{u}(\mathbf{r},\tau )\right] -i%
\mathbf{q}\cdot \mathbf{r}}\right] .  \label{1_19}
\end{equation}%
This differs from our approximation in Eq. (\ref{1_17}) essentially via the
last term which captures the short wavelength oscillations in the charge
density and allows pinning by impurities. In employing Eq. \ref{1_19}, since
only the last term can actually lead to pinning \cite{GLD95}, we will
drop the first two terms upon substitution into Eq. \ref{1_15}. Moreover, in
the reciprocal lattice sum we retain only the smallest non-trivial
wavevectors, so that in what follows (unless otherwise specified) $\sum_{%
\mathbf{K}\neq 0}$ really means sum over $\mathbf{K}=\left( \pm
K_{x0},0\right) ,\left( 0,\pm K_{y0}\right) ,\left( \pm K_{x0},\pm
K_{y0}\right) $, where $K_{x0}=2\pi /a_{x}$. These simplifications, we will
see, allow us to compute the Green's function in a relatively
straightforward manner while retaining the essential physics of pinning so
that our results are qualitatively correct. The major effect of these
approximations is to replace the soft cutoff in wavevector that would enter
through the form factor with a hard one in the reciprocal lattice sum. With
these approximations, the impurity action with which we work is 
\begin{equation}
S_{\mathrm{imp}}^{\prime }=n_{0}\int d\mathbf{r}\,d\tau \,V\left( \mathbf{r}%
\right) \,\sum_{\mathbf{K}\neq 0}e^{i\mathbf{K}\cdot \left[ \mathbf{r}-%
\mathbf{u}(\mathbf{r},\tau )\right] },  \label{impurityaction}
\end{equation}
where $n_0=1/a_xa_y$.

\subsection{Review of the RG analysis}

Before proceeding with our replica analysis, we review the highlights of the
perturbative RG analysis previously undertaken by some of us \cite{YFC} to
set the stage for our expectations for the results. In the RG approach, one
performs momentum shell integrals for large (absolute values of) frequency
and $q_{y}$, rescales the lengths and times to keep the cutoffs fixed, and
then examines how the parameters of the theory evolve under this
transformation. The power of this approach is that it may be carried out
perturbatively in the disorder, allowing one to avoid the subtleties that
arise from the employment of replicas or other methods needed to handle
disorder averages when $V\left( \mathbf{r}\right) $ remains in the exponent.

Another useful feature of the RG approach is that it allows one to look at
the contributions to the impurity action individually. Specifically, one
must modify Eq. (\ref{impurityaction}) to read 
\begin{equation}
S_{\mathrm{imp}}^{\prime }=n_{0}\int d\mathbf{r}\,d\tau \,V\left( \mathbf{r}%
\right) \,\sum_{\mathbf{K}\neq 0}\Delta _{\mathbf{K}}\left( \ell \right) e^{i%
\mathbf{K}\cdot \left[ \mathbf{r}-\mathbf{u}\left( \mathbf{r},\tau \right) %
\right] },  \label{RGflow}
\end{equation}%
where $l$ is the standard scaling variable and 
$\Delta _{\mathbf{K}}\left( \ell =0\right) =1$. The behavior of $\Delta
_{\mathbf{K}}\left( \ell \right) $ is different depending on whether $%
\mathbf{K}$ is parallel or perpendicular to the stripes. For $\mathbf{K}$ 
\textit{parallel} to the stripes, one finds 
\begin{equation}
{\frac{d\Delta _{\mathbf{K}}\left( \ell \right) }{d\ell }}=\left( {\frac{%
1-\gamma _{\mathbf{K}}}{2}}\right) \Delta _{\mathbf{K}},
\end{equation}%
with $\gamma _{\mathbf{K}}$ increasing for increasing $K$, and taking the
value 
\begin{equation}
\gamma _{\mathbf{K}}={\frac{a_{x}l_{B}^{2}}{a_{y}}}\sum_{q_{x}}d{q_{x}}{%
\frac{d_{xx}\left( q_{x}\right) }{\sqrt{d_{xx}\left( q_{x}\right)
d_{yy}\left( q_{x}\right) -d_{xy}^{2}\left( q_{x}\right) }}}-2,
\label{gammaRG}
\end{equation}%
for $\K=K_{y0}\hat{y}$, i.e., for the shortest wavevector parallel to the
stripes. The form of Eq.~(\ref{RGflow}) indicates that the stripes can undergo
a quantum phase transition, from one in which they are pinned for motion
parallel to the stripes ($\Delta _{\mathbf{K}=(2\pi /a_{y})\hat{y}}$
relevant) to one in which they are unpinned ($\Delta _{\mathbf{K}=\left(
2\pi /a_{y}\right) \hat{y}}$ irrelevant) and free to slide. As can be seen
from Eq.~(\ref{gammaRG}), which state the system ends up in depends in detail
on the elastic stiffness of the stripes. For the $N=3$ Landau level, using
the same matching procedure as we described above, it was found that the
stripes undergo a quantum depinning transition around $\Delta \nu \approx
0.43$ for very weak disorder, with the unpinned state occurring for the
larger values of $\Delta \nu $. From the form of Eq.~(\ref{RGflow}), one can see
the depinning occurs via a KT transition \cite{YFC}.

The RG analysis is more complicated if ${\mathbf{K}}$ is perpendicular to
the stripes. In this case, for any ${\mathbf{K}}=(K_{x},0)$, the free energy 
$F\equiv -\ln \int {\mathcal{D}}{\mathbf{u}}\exp \left(
-S_{0}-S_{imp}^{\prime }\right) $ diverges at low temperatures as $T^{-2/5}$
for any $\Delta \nu $. This indicates that pinning \emph{perpendicular} to
the stripes is always relevant. Our interpretation of this is that the
stripes will be trapped in channels; however, they are still free to move
along the channels so that this would not spoil the phase transition
described above.

The perturbative RG thus leads us to expect a quantum phase transition from
a pinned to an unpinned state as $\Delta\nu$ increases towards 1/2. We will
see that the replica analysis discussed below bears out this expectation,
and gives results very much in harmony with those of the perturbative RG.

Before closing this section, we believe it is important to point out that
different methods for estimating the dynamical matrix $D_{\alpha \beta }$
will lead to different values of $\gamma _{\mathbf{K}}$, and may ultimately
lead to different conclusions regarding whether there is a depinning
transition in this system. Specifically, calculations based on edge state
models for the low energy states of the stripes \cite{edgestatemodel,OC} lead
to estimates in which the stripes are always in the pinned phase for all $%
\Delta \nu\neq 1/2$. This difference does \textit{not} come as a result of a
fundamental difference in the assumed degrees of freedom for the underlying
low-energy model; indeed, one may show the edge state and disordered solid
models can be mapped onto one another \cite{edgestatemodel,Fogler1}. The
difference arises purely as a result of the different estimates one arrives
at for the dynamical matrix using the two different approaches.

A convincing argument has been made \cite{edgestatemodel} 
in the context of the edge state model
that the stripes should be in the pinned state provided the system preserves
particle-hole symmetry at $\Delta\nu=1/2$. This is not the situation for the
model we have adopted: by modeling the stripes as quantum disordered
crystals, we assume the system is isomorphic to one in which the system is
composed of point particles, which does not have this symmetry. This is
natural for our starting point, the modulated stripe HF ground state. These
states are highly reminiscent of a collection of electrons in wavepackets,
and it is natural to suppose the low-energy fluctuations will consist of
displacements of these wavepackets. Moreover, the HF groundstates from which
we start \textit{spontaneously} breaks particle-hole symmetry at $%
\Delta\nu=1/2$, arriving at a lower energy state than the uniform,
particle-hole symmetric one. Although the sliding fluctuations modify the
density to one where the particle-hole symmetry breaking may not be
immediately apparent, one does expect to see the broken symmetry in pair
correlation functions. Since our estimates of the dynamical matrix elements
are taken from the density-density response function, which is closely
related to the pair correlation function, it is not surprising that our
final result does not respect the limit set by particle-hole symmetry.

An interesting aspect of our approach is that it predicts, in the clean
limit, that there will be \textit{two} smectic states, a particle-like one,
and a hole-like one, at $\Delta\nu=1/2$. The transition between them as a
function of $\Delta\nu$ will presumably be first order. While a direct
experimental confirmation of this is difficult, the predictions we make in
the present study -- a depinning transition, and a dynamical conductivity
whose form is characteristic of a depinned state -- offer a falsifiable test
of whether the QH smectic actually breaks particle-hole symmetry:
should experiments show that the dynamical conductivity unambiguously
displays behavior associated with the depinned state, then it is most likely
that the QH smectic indeed spontaneously breaks particle-hole
symmetry at $\Delta\nu=1/2$.

\section{Beyond perturbation theory: Replicas and the GVM}

When a perturbation is relevant in an RG analysis, it is necessary to
develop some method for approximating the action to which the system is
flowing in order to compute properties of the system. For a pinned elastic
system, replicas combined with the Gaussian variational method (GVM) make
this possible. In this section, we briefly introduce this method, and go on
to discuss some aspects of its application to the stripe system. A fuller
discussion may be found in Refs. \onlinecite{MP,GLD96,GLD95,GO}.

\subsection{Basic equations}

The fundamental idea of the GVM is to replace a
complicated action $S$ with a variational action $S_{\mathrm{var}}$ that 
is \textit{quadratic}, with coefficients chosen to best match the original
problem. This is accomplished by minimizing a free energy \cite{GO} 
\begin{equation}
F_{\mathrm{var}}=F_{0}+T\left[ \left\langle S\right\rangle _{S_{\mathrm{var}%
}}-\left\langle S_{\mathrm{var}}\right\rangle _{S_{\mathrm{var}}}\right] ,
\label{var_prin}
\end{equation}%
where $S_{\mathrm{var}}$ is the quadratic variational action, $F_{0}$ is the
free energy associated with that action, and here $\left\langle \cdots
\right\rangle _{S_{\mathrm{var}}}$ indicates a functional integral over
displacements, with $S_{\mathrm{var}}$ as a weighting. For our problem, we
would like to disorder average $F_{\mathrm{var}}$, a difficult task because
the disorder potential $V$ enters $F_{\mathrm{var}}$ in a complicated and
analytically intractable way. A standard method for dealing with this is the
replica trick \cite{dotsenko}, in which one creates $n$ copies of the
original action, computes the replicated partition function $Z^{n}$ , and
then takes the $n\rightarrow 0$ limit. The identity $F=\lim_{n\rightarrow
0}\left( 1-Z^{n}\right) /n$ connects the disorder-averaged, replicated
partition function to the free energy. In practice, one first replicates
both the Gaussian variational free energy and the original action, performs
the disorder average on $Z^{n}$, and then applies Eq. \ref{var_prin} to the
resulting replicated effective action, taking the $n\rightarrow 0$ limit
only after finding the equations that come from minimizing $F_{\mathrm{var}}$%
.

Following this program, the effective replicated action after disorder
averaging is defined by 
\begin{equation}
\exp \left( -S_{\mathrm{eff}}\right) ={\frac{1}{\int DV\,P\left( V\right) }}%
\int DV\,P\left( V\right) \exp \left\{ -\sum_{a=1}^{n}\left[ S_{0}^{(a)}+S_{%
\mathrm{imp}}^{^{\prime }(a)}\right] \right\} ,
\end{equation}%
which yields 
\begin{eqnarray}
&&S_{\mathrm{eff}}=S_{0}^{(\mathrm{eff})}+S_{\mathrm{imp}}^{\mathrm{(eff)}},
\label{Seff} \\
&&S_{0}^{(\mathrm{eff})}={\frac{1}{2T}}\sum_{a=1}^{n}
\sum_{\mathbf{q},\omega _{n}}\sum_{\alpha ,\beta
=x,y}u^{a}_{\alpha }\left( \mathbf{q},\omega _{n}\right) \,G_{\alpha \beta
}^{(0)-1}\left( \mathbf{q},\omega _{n}\,\right) u^{a}_{\beta }\left( -\mathbf{q}%
,-\omega _{n}\right) ,  \label{S0eff} \\
&&S_{\mathrm{imp}}^{\mathrm{(eff)}}\simeq -v_{\mathrm{imp}%
}\sum_{a,b=1}^{n}\,\int_{0}^{1/T}d\tau _{1}\int_{0}^{1/T}d\tau _{2}\,\int d%
\mathbf{r}\sum_{\mathbf{K}\neq 0}\cos \left[ \mathbf{K}\cdot \left[ \mathbf{u%
}^{a}(\mathbf{r},\tau _{1})-\mathbf{u}^{b}(\mathbf{r},\tau _{2})\right] %
\right] ,  \label{Simpeff}
\end{eqnarray}%
where $v_{\mathrm{imp}}=V^2_0 a^2_x a^2_y n^2_0$, and 
$a,b$ are replica indices that run from $1$ to $n$. In obtaining the
last line of Eq.~(\ref{Simpeff}) we have neglected some rapidly oscillating
terms.

In the pure limit the action is diagonal in the replica indices. Disorder
averaging introduces coupling among the replicas through the impurity
coupling $S_{\mathrm{imp}}^{\mathrm{(eff)}}$ in Eq.~(\ref{Simpeff}). This
coupling is non-Gaussian, so we next apply the GVM. We introduce the
Gaussian variational action $S_{\mathrm{var}}$ which takes the form 
\begin{equation}
S_{\mathrm{var}}={\frac{1}{2T}}\sum_{\mathbf{q},\omega _{n}}u_{\alpha
}^{a}\left( \mathbf{q},\omega _{n}\right) \,\left( G^{-1}\right) _{\alpha
\beta }^{ab}\left( \mathbf{q},\omega _{n}\right) \,u_{\beta }^{b}\left( -%
\mathbf{q},-\omega _{n}\right) ,
\end{equation}%
where $G_{\alpha \beta }^{ab}\left( \mathbf{q},\omega _{n}\right) $ is the
displacement Green's function, 
\begin{equation}
G_{\alpha \beta }^{ab}\left( \mathbf{q},\omega _{n}\right)
=\int_{0}^{1/T}d\tau \left\langle T_{\tau }u_{\alpha }^{a}\left( \mathbf{q}%
,\tau \right) u_{\beta }^{b}\left( -\mathbf{q},0\right) \right\rangle _{S_{%
\mathrm{var}}}.
\end{equation}%
This quantity is to be determined through minimization of the free energy.
It is convenient to write it in terms of the bare Green's function via 
\begin{equation}
\left( G^{-1}\right) _{\alpha \beta }^{ab}\left( \mathbf{q},\omega
_{n}\right) =G_{\alpha \beta }^{(0)-1}\left( \mathbf{q},\omega _{n}\right)
\,\delta _{ab}-\zeta _{\alpha \beta }^{ab}\left( \omega _{n}\right) ,
\end{equation}%
where $\zeta _{\alpha \beta }^{ab}\left( \omega _{n}\right) $ is the
element of the variational self-energy matrix $\hat{\zeta}$ (here and hereafter
the ``hat'' indicates that the quantity is a $2\times 2$ matrix). Note that 
there is 
no $\mathbf{q}$ dependence in $\hat{\zeta}$ because we have chosen our impurity
action to be local in space; this will become clear when we find the saddle point
equations below. Note also the obvious symmetries $G^{ab}=G^{ba}$ and $\zeta
^{ab}=\zeta ^{ba}$.

Substituting $S=S_{\mathrm{eff}}$ into Eq.~(\ref{var_prin}) and performing the
functional integrals, one finds 
\begin{equation}
F_{\mathrm{var}}=F_{0}+T\left[ \left\langle S_{0}^{\mathrm{(eff)}%
}\right\rangle _{S_{\mathrm{var}}}+\left\langle S_{\mathrm{imp}}^{\mathrm{%
(eff)}}\right\rangle _{S_{\mathrm{var}}}-\left\langle S_{\mathrm{var}%
}\right\rangle _{S_{\mathrm{var}}}\right] ,
\end{equation}%
where 
\begin{eqnarray}
&&F_{0}=-{\frac{1}{2}}T\,\mathrm{Tr}\,\mathrm{ln}\,\widehat{G}+\mathrm{const.%
},  \label{F0} \\
&&\left\langle S_{0}^{\mathrm{(eff)}}-S_{\mathrm{var}}\right\rangle _{S_{%
\mathrm{var}}}={\frac{1}{2}}\sum_{\mathbf{q},\omega
_{n}}\sum_{a,b=1}^{n}\,\sum_{\alpha ,\beta =x,y}\left[ G_{\alpha \beta
}^{(0)-1}\left( \mathbf{q},\omega _{n}\right) \delta _{ab}-(G^{-1})_{\alpha
\beta }^{ab}\left( \mathbf{q},\omega _{n}\right) \right] G_{\alpha \beta
}^{ba}(\mathbf{q},\omega _{n}), \\
&&\left\langle S_{\mathrm{imp}}^{\mathrm{(eff)}}\right\rangle _{S_{\mathrm{var}%
}} =-{\frac{v_{\mathrm{imp}}}{T}}\sum_{a,b=1}^{n}\sum_{\mathbf{K}\neq
0}\int_{0}^{1/T}d\tau \exp \left[ -{\frac{1}{2}}\sum_{\alpha \beta
}K_{\alpha }K_{\beta }\,B_{\alpha \beta }^{ab}\left( \tau \right) \right] ,
\end{eqnarray}%
with 
\begin{equation}
B_{\alpha \beta }^{ab}(\tau )=\left\langle T_{\tau }[u_{\alpha }^{a}(%
\mathbf{r},\tau )-u_{\beta }^{b}(\mathbf{r},0)]^{2}\right\rangle _{S_{%
\mathrm{var}}}=T\sum_{\mathbf{q},\omega _{n}}\left[ G_{\alpha \beta }^{aa}(%
\mathbf{q},\omega _{n})+G_{\alpha \beta }^{bb}(\mathbf{q},\omega _{n})-2\cos
(\omega _{n}\tau )G_{\alpha \beta }^{ab}(\mathbf{q},\omega _{n})\right] .
\label{Bab}
\end{equation}

\subsection{Saddle point equations}

Equation (\ref{var_prin}) next needs to be extremized, which is accomplished by
taking derivatives with respect to the matrix elements of $G$, $\partial F_{%
\mathrm{var}}/\partial \hat{G}=0$. The resulting saddle point equations
(SPE's) are most easily expressed in terms of the self-energy matrix as \cite%
{GLD96,GLD95} 
\begin{eqnarray}
&&\zeta _{\alpha \beta }^{aa}(\omega _{n})=4v_{\mathrm{imp}%
}\int_{0}^{1/T}d\tau \,\left\{ \left( 1-\cos \omega _{n}\tau \right)
\,V_{\alpha \beta }^{\prime }\left[ B^{aa}(\tau )\right] +\sum_{b\neq
a}V_{\alpha \beta }^{\prime }\left[ B^{ab}(\tau )\right] \right\} ,
\label{zetaaa} \\
&&\zeta _{\alpha \beta }^{a(b\neq a)}(\omega _{n})=-4v_{\mathrm{imp}%
}\int_{0}^{1/T}d\tau \,\cos \omega _{n}\tau \,V_{\alpha \beta }^{\prime }%
\left[ B^{ab}(\tau) \right],  \label{zetaab}
\end{eqnarray}%
where 
\begin{equation}
V_{\alpha \beta }^{\prime }\left[ B^{ab}(\tau )\right] =\sum_{\mathbf{K}\neq
0}K_{\alpha }K_{\beta }\,\exp \left[ -{\frac{1}{2}}\sum_{\mu \nu =x,y}K_{\mu
}K_{\nu }B_{\mu \nu }^{ab}(\tau )\right] .
\end{equation}

It is apparent at this point that the self-energy has no $\mathbf{q}$
dependence. Moreover, if we assume that reflection symmetry for the stripe
system is not spontaneously broken after disorder averaging, it is clear
that the solutions of interest to Eqs.~(\ref{zetaaa}) and (\ref{zetaab}) will
satisfy $\zeta _{xy}^{ab}=0$. Our task will be to find the self-energy
matrix elements that are diagonal in the spatial indices.

It is now convenient to take $n\rightarrow 0$ limit. In doing so, the
replica indices are taken to be continuous rather than integral, and they
are taken from running from $1$ to $n$ to running from $1$ to $0$. An
important aspect of taking this limit is that one \textit{assumes} the
self-energy and Green's function matrices may be written in a
\textquotedblleft hierarchical form\textquotedblright\ \cite{MPV,dotsenko}.
In the limit $n\rightarrow 0$ such matrices are characterized by diagonal
and off-diagonal terms, which may be written as 
\begin{eqnarray}
&&\zeta _{\alpha \alpha }^{aa}\rightarrow \tilde{\zeta}_{\alpha }, \\
&&\zeta _{\alpha \alpha }^{ab(\neq a)}\rightarrow \zeta _{\alpha }(u),\;\;\;%
\mathrm{for}\;\;0\leq u\leq 1.
\end{eqnarray}%
Similarly, $G_{\alpha \beta }^{aa}\rightarrow \widetilde{G}_{\alpha \beta }$%
, $G_{\alpha \beta }^{ab(\neq a)}\rightarrow G_{\alpha \beta }(u)$ ($0\leq
u\leq 1$). Since the disorder potential $V(\mathbf{r})$ is time independent,
a further simplification one finds is that the off-diagonal replica
components $\zeta _{\alpha \alpha }^{ab(\neq a)}$ and $G_{\alpha \beta
}^{ab(\neq a)}$ are $\tau $ independent, \cite{GLD96,GLD95,CGLD} so
that $\widehat{G}(\mathbf{q},\omega _{n},u)$ and 
$\hat{\zeta}(\mathbf{q},\omega _{n},u)$ are different than zero only for $%
\omega _{n}=0$: 
\begin{eqnarray}
&&\widehat{G}(\mathbf{q},\omega _{n},u)=\widehat{G}(\mathbf{q},u)\,\delta
_{\omega _{n},0},  \label{nown1} \\
&&\hat{\zeta}(\omega _{n},u)=\hat{\zeta}(u)\,\delta _{\omega _{n},0}.
\label{nown2}
\end{eqnarray}%
The SPE's (\ref{zetaaa}) and (\ref{zetaab}) now may be written as 
\begin{eqnarray}
&&\tilde{\zeta}_{\alpha }(\omega _{n})=\int_{0}^{1}du\,\zeta _{\alpha
}(u)+4v_{\mathrm{imp}}\int_{0}^{1/T}d\tau \left( 1-\cos \left( \omega
_{n}\tau \right) \right) \,V_{\alpha \alpha }^{\prime }\left[ \widetilde{B}%
(\tau )\right] ,  \label{zetatilde} \\
&&\zeta _{\alpha }(u)=-{\frac{4v_{\mathrm{imp}}}{T}}V_{\alpha \alpha
}^{\prime }\left[ B(u)\right] ,  \label{zetau}
\end{eqnarray}%
where, from Eq.~(\ref{Bab}), 
\begin{eqnarray}
&&\widetilde{B}_{\mu \mu }(\tau )=2T\sum_{\mathbf{q},\omega _{n}}\left(
1-\cos (\omega _{n}\tau )\right) \widetilde{G}_{\mu \mu }(\mathbf{q},\omega
_{n}),  \label{Btilde} \\
&&B_{\mu \mu }(u)=2T\sum_{\mathbf{q}}\left\{ \left[ \sum_{\omega _{n}}%
\widetilde{G}_{\mu \mu }(\mathbf{q},\omega _{n})\right] -G_{\mu \mu }(%
\mathbf{q},u)\right\}   \notag \\
&&\;\;\;\;\;\;\;\;=2T\sum_{\mathbf{q}}\left\{ \left[ \widetilde{G}_{\mu \mu
}(\mathbf{q},\omega _{n}=0)-G_{\mu \mu }(\mathbf{q},u)\right] +\sum_{\omega
_{n}\neq 0}\widetilde{G}_{\mu \mu }(\mathbf{q},\omega _{n})\right\} .
\label{Bu}
\end{eqnarray}%
Note that Eq.~(\ref{zetatilde}) also gives us 
\begin{equation}
\tilde{\zeta}_{\alpha }(\omega _{n}=0)=\int_{0}^{1}du\zeta _{\alpha }(u).
\end{equation}

To solve the Eqs.~(\ref{zetatilde}) and (\ref{zetau}) we must know the
relation between $\widetilde{G}(\omega _{n})$, $G(u)$ and $\tilde{\zeta}%
(\omega _{n})$ and $\zeta (u)$. Eqs.~(\ref{nown1}) and (\ref{nown2})
indicate that 
\begin{equation}
\widetilde{G}_{\mu \mu }(\mathbf{q},\omega _{n}\neq 0)=\left[ \widehat{G}%
^{(0)-1}(\mathbf{q},\omega _{n})-\hat{\tilde{\zeta}}(\omega _{n})\right]
_{\mu \mu }^{-1}.  \label{tildeG}
\end{equation}%
The quantities $\widehat{\widetilde{G}}(\mathbf{q},\omega _{n}=0)$ and $\hat{%
G}(u)$ are related to $\hat{\tilde{\zeta}}(\omega _{n}=0)$ and $\hat{\zeta}%
(u)$ through inversion rules that generalize the inversion of an $n\times n$
hierarchical matrix to the $n\rightarrow 0$ limit. The inversion rules for a
simple hierarchical matrix are well-known \cite{MP,dotsenko}, and their
generalization to a situation in which the elements of the hierarchical
matrix are proportional to the unit matrix -- which would be the case for
our matrices if the elastic system were isotropic -- is trivial. However, in
our case the entries of the hierarchical matrix are $2\times 2$ matrices
with a non-trivial structure. Moreover, the perturbative RG indicates we
should expect the pinning properties perpendicular and parallel to the
stripes to be different, and we need to generalize the inversion rules to
allow for this possibility. With some work, the most general inversion rules
for our situation can be derived analytically, and we present this
derivation in Appendix~\ref{inversionrules}. According to Eq.~(\ref%
{inversion3}), the Green's functions are related to the self-energy by 
\begin{equation}
\widehat{\widetilde{G}}(\mathbf{q},\omega _{n}=0)-\widehat{G}(\mathbf{q},u)=%
\left[ \widehat{D}(\mathbf{q})-\hat{\tilde{\zeta}}(\omega _{n}=0)+\hat{\zeta}%
(u)\right] ^{-1}+\int_{u}^{1}dv\left[ \widehat{D}(\mathbf{q})+\left[ \hat{%
\zeta}\right] (v)\right] ^{-1}\cdot \hat{\zeta}^{\prime }(v)\cdot \left[ 
\widehat{D}(\mathbf{q})+\left[ \hat{\zeta}\right] (v)\right] ^{-1},
\label{Ginversion}
\end{equation}%
where $\hat{\zeta}^{\prime }(v)=d\hat{\zeta}(v)/dv$, and 
\begin{equation}
\left[ \hat{\zeta}\right] (u)=u\,\hat{\zeta}(u)-\int_{0}^{u}dv\,\hat{\zeta}%
(v).  \label{[zetau]}
\end{equation}%

Once we have obtained the self-energy, we can compute the finite-frequency
conductivities in Eq.~(\ref{conductivity0}) by analytically continuing to
real frequency in Eq.~(\ref{tildeG}), so that 
\begin{eqnarray}
\widetilde{G}_{\mu \mu }^{\mathrm{ret}}(\mathbf{q},\omega \neq 0)&=&\left[ 
\widehat{G}_{\mathrm{ret}}^{(0)-1}(\mathbf{q},\omega )-\hat{\tilde{\zeta}}^{%
\mathrm{ret}}(\omega )\right] _{\mu \mu }^{-1}.  \label{Gomega}
\end{eqnarray}%
Inserting Eq.~(\ref{Gomega}) into Eq.~(\ref{conductivity0}) we arrive at the
longitudinal conductivity 
\begin{equation}
\sigma _{\alpha \alpha }(\omega )={\frac{e^{2}}{a_{x}a_{y}}}{\frac{i\omega 
\tilde{\zeta}_{\bar{\alpha}}^{\mathrm{ret}}(\omega )}{\tilde{\zeta}_{x}^{%
\mathrm{ret}}(\omega )\tilde{\zeta}_{y}^{\mathrm{ret}}(\omega )-\omega
^{2}/l_{B}^{4}},}  \label{longitudinalconductivity}
\end{equation}%
where $\bar{\alpha}=y\,(x)$ for $\alpha =x\,(y)$.

To obtain $\tilde{\zeta}_{\alpha }^{\mathrm{ret}}(\omega )$, we analytically
continue Eq.~(\ref{zetatilde}). As shown in Appendix~\ref%
{Appendixanalycon}, this results in the equation (for $T=0$) 
\begin{equation}
\tilde{\zeta}_{\alpha }^{\mathrm{ret}}(\omega )=e_{\alpha }-4v_{\mathrm{imp}%
}\sum_{\mathbf{K}\neq 0}K_{\alpha }^{2}\int_{0}^{\infty }dt\;(e^{i\omega
t}-1)\mathrm{Im}\left[ I(t,K_{x})I\left( t,K_{y}\right) \right] ,
\label{SPEs}
\end{equation}%
where 
\begin{eqnarray}
&&I\left( t,K_{\mu }\right) =\mathrm{\exp }\left[ -{\frac{K_{\mu }^{2}}{\pi }%
}\int_{0}^{\infty }df\,A_{\mu }(f)\left( 1-e^{itf}\right) \right] ,
\label{Ialpha} \\
&&A_{\mu }(f)=\sum_{q}\mathrm{Im}\left[ \widetilde{G}_{\mu \mu }^{\mathrm{ret%
}}(\mathbf{q},f)\right] , \label{spectralfunction}\\
&&e_{\alpha }=\tilde{\zeta}_{\alpha }^{\mathrm{ret}}(0^{+})=\int_{0}^{1}du\,%
\zeta _{\alpha }(u)-4v_{\mathrm{imp}}\sum_{\mathbf{K}\neq 0}K_{\alpha
}^{2}\int_{0}^{\infty }dt\;\mathrm{Im}\left[ I(t,K_{x})I\left(
t,K_{y}\right) \right]  \notag \\
&&\;\;\;\;\;\;\;\;\;\;\;\;\;\;\;\;-2v_{\mathrm{imp}}\sum_{\mathbf{K}\neq
0}K_{\alpha }^{2}\int_{0}^{1/T}d\tau \exp \left[ -{\frac{1}{2}}\sum_{\mu
=x,y}K_{\mu }^{2}\widetilde{B}_{\mu \mu }(\tau )\right] .  \label{e0}
\end{eqnarray}%
As we proceed with our analysis, it is helpful to keep in mind that $A_{\mu
}(f)$ is a spectral function, and that $e_{\alpha }\neq 0$ is an energy
offset that in a pinned state opens a gap in the phonon spectrum, as
discussed more fully below. Note also that Eq.~(\ref{SPEs}) indicates that $%
\tilde{\zeta}_{\alpha }^{\mathrm{ret}}(\omega )$ at each $\omega $ point
depends on the whole spectrum of $A_{\alpha }(f)$, and this will be
complications that for this analysis cannot be avoided, as is the case for
other pinned systems \cite{GLD96,GLD95}. Since quantum fluctuations play a
crucial role in this system, it is useful for us to define effective
Debye-Waller factors via 
\begin{equation}
W(\mathbf{K})={\frac{1}{\pi }}\sum_{\mu =x,y}K_{\mu }^{2}\int_{0}^{\infty
}df\,A_{\mu }(f).
\end{equation}%
These quantities are a measure of the mean square displacements in units of
the lattice constants, and when large they indicate that quantum
fluctuations cannot be ignored in computing the dynamical conductivity.
Clearly this will be the case in the vicinity of the quantum depinning
transition. On the other hand, if $W(\mathbf{K})$ are small for all $\mathbf{%
K}$, one may expand the exponential function on the right-hand side of Eq.~(%
\ref{Ialpha}) and keep only the leading order term. Eqs.~(\ref{SPEs}) then
become greatly simplified, taking the form 
\begin{equation}
\tilde{\zeta}_{\alpha }^{\mathrm{ret}}(\omega )=e_{\alpha }+2v_{\mathrm{imp}%
}\sum_{\mathbf{K}\neq 0}K_{\alpha }^{2}\sum_{\mu =x,y}K_{\mu }^{2}\sum_{%
\mathbf{q}}\left[\tilde{G}_{\mu \mu }^{\mathrm{ret}}(\mathbf{q},\omega )
- \tilde{G}_{\mu \mu }^{\mathrm{ret}}(\mathbf{q},\omega=0^+) \right ].
\label{SPEsc}
\end{equation}%
This is called the semiclassical approximation (SCA) \cite{GLD96,GLD95}, and
it presents a powerful simplification when it is valid. In particular one
sees that Eq.~(\ref{SPEsc}) is local in the frequency, so that $\tilde{\zeta}%
_{\alpha }^{\mathrm{ret}}(\omega )$ may be determined one frequency at a
time. Unfortunately, the SCA is not valid in our present problem, and we are
forced to solving the full SPE's (\ref{SPEs}) numerically. We will see
however that the solutions have several interesting properties that give
clear signatures of the depinned phase and the transition leading to it.

\subsection{Replica symmetric (RS) solution vs replica symmetry breaking
(RSB) solution}

Eq.~(\ref{SPEs}) shows that the replica diagonal self energy $\tilde{\zeta}^{%
\mathrm{ret}}_{\alpha}(\omega)$ depends on the off-diagonal terms $%
\zeta_{\alpha}(u)$ through the constants $e_\alpha$. It is instructive to
first examine the possible structure of $\zeta_{\alpha}(u)$. If $%
\zeta_{\alpha}(u)$ is a constant in $u$, the symmetry of permutation of the
replica indices is kept and the solution is ``replica symmetric'' (RS). On
the other hand, when $\zeta_{\alpha}(u)$ varies with $u$, the solution
displays replica symmetry breaking (RSB).

For many low-dimensional systems ($d\leq 2$), the appropriate solution to
the SPE's is of the RSB type. Often there is a simple \textquotedblleft
one-step RSB\textquotedblright\ solution, with $\zeta (u)$ piecewise
constant, but stepping up or down at a single point $u_{c}$ ($0<u_{c}<1$).
It follows from Eq.~(\ref{[zetau]}) that $\left[ \zeta _{\alpha }\right]
(u_{c})\neq 0$ in the RSB state. On the other hand, $\left[ \zeta _{\alpha }%
\right] (u_{c})=0$ for the RS solution.

Following Refs. \onlinecite{GLD96} and \onlinecite{GO}, one can establish a 
close relation between $\left[ \zeta _{\alpha }\right] (u_{c})$ and $e_{\alpha }$. 
By making use of $\left[ \zeta _{\alpha }\right] (u_{c})$, Eq.~(\ref{zetatilde}%
) can be rewritten as 
\begin{equation}
\tilde{\zeta}_{\alpha }(\omega _{n}\neq 0)=\left[ \zeta _{\alpha }\right]
(u_{c})-4v_{\mathrm{imp}}\int_{0}^{1/T}d\tau \left( 1-\cos \left( \omega
_{n}\tau \right) \right) \,\left\{ V_{\alpha \alpha }^{\prime }[\widetilde{B}%
(\tau )]-V_{\alpha \alpha }^{\prime }[B(u_{c})]\right\} .  \label{spe1}
\end{equation}%
Here, substituting $V_{\alpha \alpha }^{\prime }\left[ B(u_{c})\right] $
from $V_{\alpha \alpha }^{\prime }\left[ \widetilde{B}(\tau )\right] $
guarantees that as $T\rightarrow 0$ the second term of the right-hand side
of Eq.~(\ref{spe1}) vanishes at $\omega _{n}\rightarrow 0$. Comparing Eqs.~(%
\ref{spe1}) and (\ref{SPEs}) one immediately concludes that 
\begin{equation}
e_{\alpha }=-\left[ \zeta _{\alpha }\right] (u_{c}).  \label{e01}
\end{equation}%
So $e_{\alpha }=0$ in the RS state and $e_{\alpha }\neq 0$ in the RSB state.
The two constants $e_{x}$ and $e_{y}$ have significant physical meanings.
They may be regarded as a measure of the strength of pinning by the disorder
potential and are roughly speaking proportional to the gap in the low-energy
magnetophonon modes. If $e_{\alpha }=0$, the phonon spectrum is gapless at ${%
\mathbf{q}}=0$ indicating the system can slide as a whole without energy
cost, and is not pinned. Thus an RS solution is expected in the unpinned
state. If $e_{\alpha }\neq 0$ as in the RSB solution, a gap opens up in the
low-energy magnetophonon modes, uniform sliding cannot be achieved at zero
energy, and the system is pinned by disorder.

We will show in Sec.~V that for $\Delta\nu<\Delta\nu_c$, both $e_x$ and $e_y$
are nonzero and the stripes are thus fully pinned. As $\Delta\nu\rightarrow
\Delta\nu_c$, $e_y\rightarrow 0$, indicating a quantum depinning transition.
The solution to the SPE's is RS for motion along the stripes but RSB for
motion perpendicular to the stripes. We call this type of solution to the
SPE's a \textit{partial} RSB state. The detailed behavior of the
system in this state will be explored in Sec.~VI.

\section{Results for pinned state: RSB solution}

We begin by examining solutions of the SPE's for which the QH stripes are
fully pinned by disorder. According to the RG result reviewed in Sec.~III~B,
this corresponds to $\Delta\nu <\Delta\nu_c$ in which the disorder is
relevant. In this case both $e_x$ and $e_y$ are nonzero. We begin by
discussing constraints on $e_x$ and $e_y$ which determine their values,
allowing us to solve the SPE's without explicitly solving for $%
\zeta_{\alpha}(u)$ and $\tilde{\zeta}_{\alpha}$. We then present numerical
results for the conductivity.

\subsection{Two constraints for $e_{x}$ and $e_{y}$}

The SPE's (\ref{SPEs}) become a set of closed equations if $e_{x}$ and $%
e_{y} $ are known. Formally, $e_{x}$ and $e_{y}$ need to be determined
self-consistently by solving Eq.~(\ref{zetau}). The SPE's in fact have a
family of solutions (parameterized by $u_{c}$), and determining which is
best generically would be determined by minimization of the free energy. In
the case of spatial dimension $d>2$, $u_{c}$ determined this way leads to Re$%
\left[ \sigma (\omega )\right] \sim \omega ^{2}$ at small $\omega $. This is
consistent with arguments by Mott as well as some exact solutions \cite%
{exactsolution} (up to a logarithmic correction). However, for $d\leq 2$,
this approach can yield an unphysical result in which, in the pinned state,
the conductivity shows a true gap: Re$\left[ \sigma \left( \omega \right) %
\right] $ vanishes below some gap frequency. Alternatively, one may \textit{%
impose} the condition Re$\left[ \sigma (\omega )\right] \sim \omega ^{2}$ at
small $\omega $. It is known that the doing so generates an equation that
may be understood as imposing a marginal stability on the so-called replicon
mode \cite{GLD96}. Although this point is not fully understood, it is a
common procedure that leads to physically reasonable results, and we will
adopt it by imposing the condition Re$\left[ \sigma _{\alpha \alpha }(\omega
)\right] \sim \omega ^{2}$ at small $\omega $ in the pinned state. From Eq.~(%
\ref{longitudinalconductivity}), this is equivalent to Im$\left[ \zeta
_{\alpha }^{\mathrm{ret}}(\omega )\right] \sim \omega $. Note that this
guarantees the magnetophonon mode density of state vanishes at zero
frequency, as one should expect for a pinned system.

To obtain the explicit condition leading to Im$\left[ \zeta _{\alpha }^{%
\mathrm{ret}}(\omega )\right] \sim \omega $, we expand the SPE's, 
Eqs.~(\ref{SPEs}), for small-$\omega $. The integral over $t$ now is dominated 
by the large $t$ region. Therefore, the term $\int_{0}^{\infty }df\,A_{\mu
}(f)e^{ift}$ in the argument of the exponential function in Eq.~(\ref{Ialpha}%
) must be small due to the rapidly oscillating nature of $e^{ift}$, leading
to 
\begin{equation}
I(t,K_{x})I(t,K_{y})\simeq e^{-W(\mathbf{K})}\left[ 1+{\frac{1}{\pi }}%
\sum_{\mu =x,y}K_{\mu }^{2}\int_{0}^{\infty }df\,A_{\mu }(f)e^{itf}\right] .
\end{equation}%
The SPE's (\ref{SPEs}) at small $\omega $ become 
\begin{equation}
\tilde{\zeta}_{\alpha }^{\mathrm{ret}}(\omega )\simeq e_{\alpha }+2\sum_{%
\mathbf{K}\neq 0}v(\mathbf{K})\,K_{\alpha }^{2}\sum_{\mu =x,y}K_{\mu
}^{2}\sum_{q} \left[\tilde{G}_{\mu \mu }^{\mathrm{ret}}(\mathbf{q},\omega )
- \tilde{G}_{\mu \mu }^{\mathrm{ret}}(\mathbf{q},\omega=0^+)\right ],
\label{smallw}
\end{equation}%
where $v(\mathbf{K})=v_{\mathrm{imp}}e^{-W(\mathbf{K})}$. This is very
similar to the SPEs (\ref{SPEsc}) within the SCA except that $v_{\mathrm{imp}%
}$ in Eq.~(\ref{SPEsc}) is now replaced by $v(\mathbf{K})$. Apparently, when 
$W(\mathbf{K})\ll 1$ the semiclassical approximation is valid, and Eq.~(\ref%
{smallw}) reduces to Eq.~(\ref{SPEsc}).

At small $\omega $, we write 
\begin{eqnarray}
&&\mathrm{Re}\left[ \tilde{\zeta}_{\alpha }^{\mathrm{ret}}\right] \simeq
e_{\alpha },  \label{rezetasmallwpin} \\
&&\mathrm{Im}\left[ \tilde{\zeta}_{\alpha }^{\mathrm{ret}}\right] \simeq
\beta _{\alpha }\omega ,  \label{imzetasmallwpin}
\end{eqnarray}%
and correspondingly $\sum_{\mathbf{q}}\tilde{G}_{\mu \mu }^{\mathrm{ret}}(%
\mathbf{q},\omega )\simeq G_{\mu 0}+\sum_{\alpha =x,y}g_{\mu \alpha }\beta
_{\alpha }\omega $. The condition for nonvanishing $\beta _{\alpha }$ from
Eq.~(\ref{smallw}) becomes 
\begin{equation}
(U_{xx}-1)(U_{yy}-1)-U_{yx}U_{xy}=0,  \label{constraint1}
\end{equation}%
where 
\begin{equation}
U_{\mu \nu }=2\sum_{\mathbf{K}\neq 0}K_{\nu }^{2}\,v(\mathbf{K})\sum_{\alpha
=x,y}K_{\alpha }^{2}g_{\alpha \mu }.  \label{U}
\end{equation}%
Eq.~(\ref{constraint1}) is our first constraint for $e_{x}$ and $e_{y}$.

\vspace{0.4cm} 
The second constraint follows from the assumption of a \emph{%
one-step} RSB solution in which $\zeta _{\alpha }(u<u_{c})=0$. Eqs.~(\ref%
{e01}) and (\ref{[zetau]}) immediately yield 
\begin{equation}
e_{\alpha }=-u_{c}\zeta _{\alpha }\left( u_{c}\right) .  \label{ea}
\end{equation}%
Inserting Eq.~(\ref{ea}) into Eq.~(\ref{Ginversion}) and noting $\zeta
_{\alpha }^{\prime }(v)=0$ for $v\geq u_{c}$ we get 
\begin{equation}
\widehat{\widetilde{G}}(\mathbf{q},n=0)-\widehat{G}(\mathbf{q},u_{c})=\left[ 
\widehat{D}(\mathbf{q})+\hat{e}\right] ^{-1},  \label{dG}
\end{equation}%
where the elements of the matrix $\hat{e}$ are $e_{\alpha \beta }=e_{\alpha
}\delta _{\alpha \beta }$. Substituting Eq.~(\ref{dG}) in (\ref{Bu}) and
making use of $\widetilde{G}_{\mu \mu }(\mathbf{q},\omega _{n}\rightarrow 0)=%
\left[ \widehat{D}(\mathbf{q})+\hat{e}\right] _{\mu \mu }^{-1}$ results in
the equation 
\begin{equation}
B_{\mu \mu }(u_{c})={\frac{2}{\pi }}\int_{0}^{\infty }dfA_{\mu }(f),
\label{Buc}
\end{equation}%
which can be inserted in Eq.~(\ref{zetau}) to give 
\begin{equation}
\zeta _{\alpha }(u_{c})={\frac{4v_{imp}}{T}}\sum_{\mathbf{K}\neq 0}K_{\alpha
}^{2}e^{-W(\mathbf{K})}.  \label{zetauc}
\end{equation}%
Eqs.~(\ref{zetauc}) and (\ref{ea}) lead to the ratio 
\begin{equation}
{\frac{e_{y}}{e_{x}}}={\frac{\sum_{\mathbf{K}\neq 0}K_{y}^{2}\,e^{-W(\mathbf{%
K})}}{\sum_{\mathbf{K}\neq 0}K_{x}^{2}\,e^{-W(\mathbf{K})}}}.
\label{constraint2}
\end{equation}%
This shows that the pinning of the stripes will generically be anisotropic,
and serves as our second constraint for $e_{\alpha }$.

The appearance of the Debye-Waller factors $W(\mathbf{K})$ in Eq.~(\ref%
{constraint2}) has a significant impact: they are responsible for the change
of behavior in $\tilde{\zeta}^{\mathrm{ret}}_{\alpha}(\omega)$ across the
depinning transition. As we shall see, whenever $K_y\neq 0$, $W(\mathbf{K})$
increases as $\Delta\nu\rightarrow \Delta\nu_c$ from below, and it
eventually diverges at $\Delta\nu_c$ leading to a suppression of $e_y$. We
will discuss this in detail below. We stress that this behavior cannot be
captured by the semiclassical approximation.

\subsection{Numerical results}

We are now in a position to solve the problem numerically. For a given pair
of $e_{x}$ and $e_{y}$, we use an iterative method to solve for $\tilde{\zeta%
}_{\alpha }^{\mathrm{ret}}(\omega )$ from the SPE's (\ref{SPEs}).
(Typically 20-30 iterations lead to a good convergence.) The computed $%
\tilde{\zeta}_{\alpha }^{\mathrm{ret}}(\omega )$ are then inserted in the
two constraint equations (\ref{constraint1}) and (\ref{constraint2}) to
generate new values of $e_{\alpha }$, and the entire process is repeated
until we reach self-consistency. We work in the $N=3$ Landau level, although
different Landau indices should give similar results. All our calculations
are obtained for a disorder level $v_{\mathrm{imp}}=0.0005e^{4}/l_{B}^{2}$.
This is likely to be somewhat larger than experimental values, but we choose
it for numerical convenience \cite{footnotedisorderlevel}. We do not expect
our results to qualitatively change for smaller disorder strengths. We note
that the bending term in Eq.~(\ref{smecticDxx}) plays an important role of
eliminating an artificial ultraviolet divergence at large $q_{y}$, but
beyond this has little effect. We choose $\kappa _{b}=2$ for all the
fillings since this leads to a relatively fast convergence of the SPE's, 
although we believe the value should be somewhat smaller (of order 1).

\begin{figure}[h]
\begin{picture}(200,200)
\leavevmode\centering\includegraphics{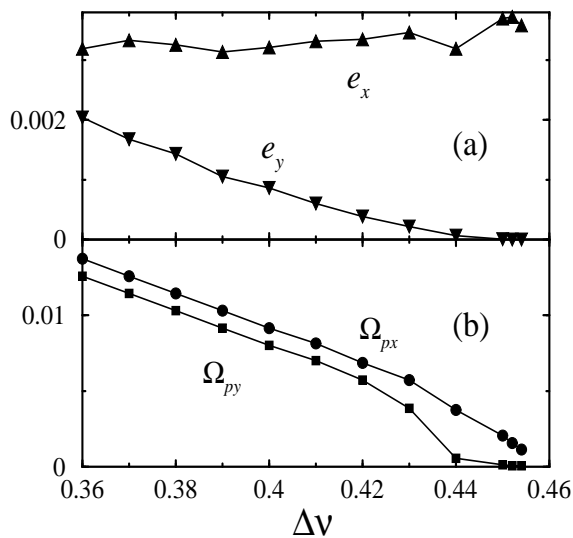}
\end{picture}
\caption{(a) Constants $e_{x}$ and $e_{y}$ in units of $e^{2}/l_{B}^{3}$,
and (b) peak positions $\Omega _{px}$ and $\Omega _{py}$ in units of $%
e^{2}/l_{B}$, as functions of $\Delta \protect\nu $ in the pinned state. }
\label{figpeak}
\end{figure}

Results for $e_x$ and $e_y$ as functions of the partial filling in the
pinned state are shown in Fig.~\ref{figpeak}~(a). The quantity $e_x$ is a
weak function of $\Delta\nu$, but $e_y$ decreases with increasing $\Delta\nu$%
, and eventually vanishes at $\Delta\nu=\Delta\nu_c\simeq 0.459$. This is
the consequence of a divergence in $W(K_x;K_y\neq 0)$ at $%
\Delta\nu=\Delta\nu_c$. Note $\Delta\nu_c$ is somewhat larger than what was
found in the perturbative RG \cite{YFC}. This is due to the non-vanishing
disorder strength; as $v_{\mathrm{imp}}$ decreases, $\Delta\nu_c$ decreases
to the value found in Ref. \onlinecite{YFC}.

\begin{figure}[h]
\begin{picture}(250,200)
\leavevmode\centering\includegraphics{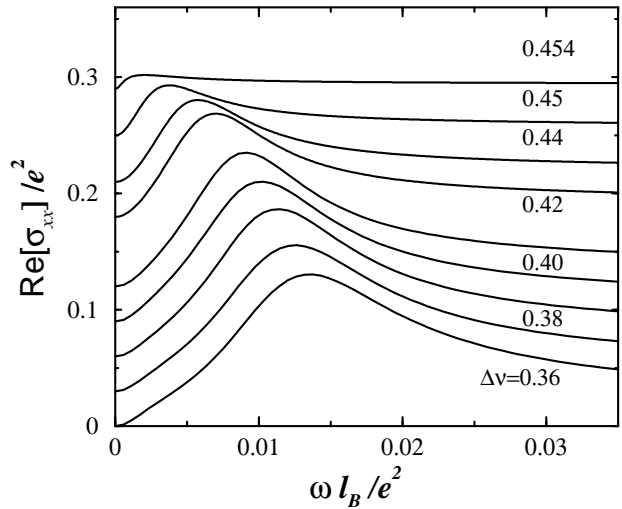}
\end{picture}
\caption{Real part of conductivity perpendicular to the stripes as a
function of frequency in the pinned (RSB) state. The disorder strength $v_{%
\mathrm{imp}}=0.0005e^{4}/l_{B}^{2}$ is used. All curves start from Re$\left[
\protect\sigma _{xx}\left( \protect\omega \right) \right] =0$ at $\protect%
\omega =0$, and curves except for $\Delta \protect\nu =0.36$ are lifted
upward for clarity. }
\label{figsgmxpin}
\end{figure}

The dynamical conductivities perpendicular to the stripes in a pinned (RSB)
phase are presented in Fig.~\ref{figsgmxpin}. For $\Delta \nu $ well below $%
\Delta \nu _{c}\approx 0.459$, Re$\left[ \sigma _{xx}\left( \omega \right) %
\right] $ has a pinning peak whose lineshape is qualitatively similar to
what is found using the SCA. \cite{CGLD}. The prominent behavior visible in
Fig.~\ref{figsgmxpin} is a monotonic decrease of the peak frequency $\Omega
_{px}$ with growing $\Delta \nu $, and its eventual collapse as the
depinning transition is approached. The peak frequency behavior is more
clearly shown in Fig.~\ref{figpeak}~(b). Notice the lineshape becomes
increasingly asymmetric as the transition is approached. Experimental
observations so far seem to be consistent with this\cite{muwave,Florida}.

\begin{figure}[h]
\begin{picture}(250,200)
\leavevmode\centering\includegraphics{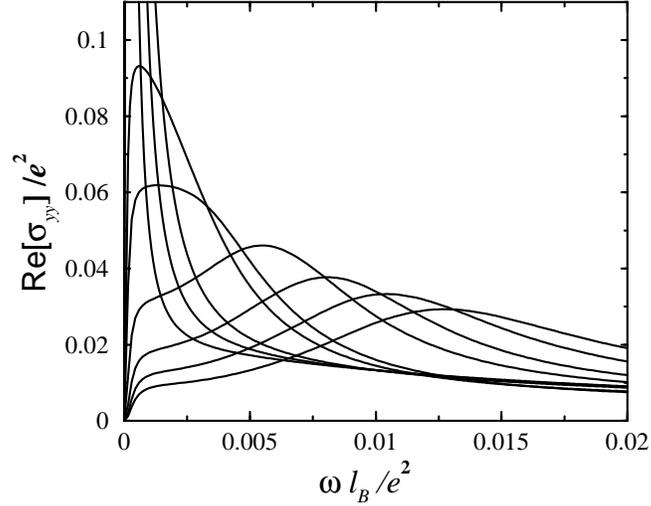}
\end{picture}
\caption{Real part of conductivity along the stripes as a function of
frequency in the pinned (RSB) state. The $v_{\mathrm{imp}}$ is the same as
in Fig.~\protect\ref{figsgmxpin}. Curves from right to left correspond to $%
\Delta \protect\nu =0.36,0.38,0.4,0.42,0.43,0.44,0.45,0.452,0.454$,
respectively. }
\label{figsgmypin}
\end{figure}

The real part of the conductivity along the stripes Re$\left[ \sigma
_{yy}\left( \omega \right) \right] $ is shown in Fig.~\ref{figsgmypin}. It
also presents a pinning peak whose frequency $\Omega _{py}$ falls down with
increasing $\Delta \nu $ as shown in Fig.~\ref{figpeak}~(b). But the
observed peak lineshape is more interesting than that of Re$\left[ \sigma
_{xx}\left( \omega \right) \right] $. Below the peak frequency $\Omega _{py}$%
, in the range $e_{y}<\omega <\Omega _{py}$ the conductivity appears to tend
toward a non-vanishing value when $\Delta \nu $ is sufficiently below $%
\Delta \nu _{c}$; only for $\omega $ well below this range does one find Re$%
\left[ \sigma _{yy}(\omega )\right] $ decreasing. The reason for this is
that the quantity $e_{y}$ turns out to be rather small [as shown in Fig.~\ref%
{figpeak}~(a)] due to a large Debye-Waller factor, and in this frequency
range the system displays a behavior similar to an incoherent metal response 
\cite{kohn}. We discuss this in more detail for the depinned (PRSB) phase
below. For $\omega \ll e_{y}$, Re$\left[ \sigma _{yy}(\omega )\right] $
vanishes quadratically with $\omega $ (not visible on the scale of Fig.~\ref%
{figsgmypin}), as required for a pinned state. As $\Delta \nu \rightarrow
\Delta \nu _{c}$, we eventually reach a situation in which $e_{y}$ and $%
\Omega _{py}$ are of similar order, in which case the pinning peak sharpens
and grows quite large. This peak continuously evolves into a $\delta $%
-function at zero frequency as the system enters into the PRSB state, so
that the transition from pinned to depinned behavior is very continuous.

Interestingly, as shown in Fig.~\ref{figpeak}, $e_{y}\rightarrow 0$ governs
the vanishing of both $\Omega _{px}$ and $\Omega _{py}$. To understand this,
we note that there are two gapless collective modes in the absence of the
magnetic field; the magnetic field mixes them into two other modes, one of
which is at a high value (order of $\hbar \omega _{c}$), leaving the other
(magnetophonon) mode as the only gapless one. It is this single mode that
responds to the electric field, albeit in an anisotropic manner in the $\hat{%
x}$ and $\hat{y}$ directions. Technically, at $\omega \sim \Omega
_{px},\Omega _{py}$, $\zeta _{\alpha }^{\mathrm{ret}}(\omega )$ obeys Eqs.~(%
\ref{rezetasmallwpin}-\ref{imzetasmallwpin}), and the longitudinal
conductivities in Eq.~(\ref{longitudinalconductivity}) become 
\begin{equation}
\mathrm{Re}\left[ \sigma _{\alpha \alpha }(\omega )\right] \simeq {\frac{%
e^{2}}{a_{x}a_{y}}}\;\omega ^{2}{\frac{e_{\bar{\alpha}}(e_{x}\beta
_{y}+e_{y}\beta _{x})+\beta _{\bar{\alpha}}\left[ \left( 1+\beta _{x}\beta
_{y}\right) \omega ^{2}-e_{x}e_{y})\right] }{\left[ \left( 1+\beta _{x}\beta
_{y}\right) \omega ^{2}-e_{x}e_{y}\right] ^{2}+\omega ^{2}\left[ e_{x}\beta
_{y}+e_{y}\beta _{x}\right] ^{2}}}.  \label{sigma}
\end{equation}%
From this we can extract 
\begin{equation}
\Omega _{px}\sim \Omega _{py}\sim \sqrt{\frac{e_{x}e_{y}}{1+\beta _{x}\beta
_{y}}}.
\end{equation}

\section{Results for depinned state: Partial RSB (PRSB) solution}

For $\Delta\nu\ge \Delta\nu_c$ the state is characterized by $e_x\ne 0$ but $%
e_y=0$. As discussed in Sec.~IV.C, this corresponds to a RSB solution for $%
\zeta_x(u)$ but a RS solution for $\zeta_y(u)$. We call this the \emph{PRSB}
state. This state has various interesting properties that we will present
below.

\subsection{Power law behavior for $\tilde{\protect\zeta}_{y}^{\mathrm{ret}}(%
\protect\omega )$}

From the results of the perturbative RG, we expect the stripes to remain
pinned for motion in the $\hat{x}$ direction even as the stripes become
depinned for motion in the $\hat{y}$ direction. We therefore assume that in
the PRSB state, the small-$\omega $ asymptotic behavior of $\tilde{\zeta}%
_{x}^{\mathrm{ret}}(\omega )$ remains the same as that in the RSB state
[Eqs.~(\ref{rezetasmallwpin}-\ref{imzetasmallwpin})], and this turns out to
yield a self-consistent solution. However, $\tilde{\zeta}_{y}^{\mathrm{ret}%
}(\omega )$ is qualitatively different in the depinned state. To see this
explicitly we examine the SPEs (\ref{SPEs}) for $\tilde{\zeta}_{y}^{\mathrm{%
ret}}(\omega )$, 
\begin{equation}
\tilde{\zeta}_{y}^{\mathrm{ret}}(\omega )=-4v_{\mathrm{imp}}\left\{
2K_{y0}^{2}\int_{0}^{\infty }dt\;\left( e^{i\omega t}-1\right) \mathrm{Im}%
\left[ I(t,K_{y0})\right] +4K_{y0}^{2}\int_{0}^{\infty }dt\;(e^{i\omega t}-1)%
\mathrm{Im} \left[I(t,K_{y0})I(t,K_{x0})\right]\right\} .  \label{SPEPRSB}
\end{equation}%
There is a self-consistent solution for this equation in which $\tilde{\zeta}%
_{y}^{\mathrm{ret}}(\omega )$ has an anomalous power law behavior at low
frequencies, 
\begin{eqnarray}
&&\mathrm{Re}\left[ \mathrm{~}\tilde{\zeta}_{y}^{\mathrm{ret}}(\omega )%
\right] \simeq \alpha _{y}\omega ^{2},  \label{rezetasmallwdepin} \\
&&\mathrm{Im~}\left[ \tilde{\zeta}_{y}^{\mathrm{ret}}(\omega )\right] \simeq
\beta _{y}\omega ^{\gamma +1}.  \label{imzetasmallwdepin}
\end{eqnarray}%
This solution is only valid when 
\begin{equation}
\gamma \geq 1,  \label{inequality}
\end{equation}%
where $\gamma $ is defined in Eq.~(\ref{gamma}) below. Note that at small $%
\omega $, $\tilde{\zeta}_{y}^{\mathrm{ret}}(\omega )$ is dominated by large-$%
t$ behavior of the integrands in Eq.~(\ref{SPEPRSB}), and this in turn
depends on small-$f$ asymptotics of $A_{\alpha }(f)\left( 1-e^{itf}\right) $
in $I(t,K_{\alpha 0})$ [see Eq.~(\ref{Ialpha})]. Making use of Eqs.~(\ref%
{rezetasmallwpin}-\ref{imzetasmallwpin}) and (\ref{rezetasmallwdepin}-\ref%
{imzetasmallwdepin}), and of the smectic form of the dynamical matrix $%
\widehat{D}(\mathbf{q})$ in Eqs.~(\ref{smecticDxx}-\ref{smecticDyy}) (with
the bending term neglected), we obtain 
\begin{equation}
A_{y}(f)\simeq {\frac{a_{x}a_{y}}{(2\pi )^{2}}}\int dq_{x}\left[
d_{xx}(q_{x})-e_{x}\right] \,\mathrm{Im}\,\left[ \int_{-\infty }^{\infty }{%
\frac{dq_{y}}{g_{x}(q_{x})q_{y}^{2}-(1+\alpha _{y})\left( f+i0^{+}\right)
^{2}}}\right] \simeq {\frac{\pi c_{v}}{f}},  \label{Ay}
\end{equation}%
where 
\begin{equation}
g_{x}(q_{x})=(1+\alpha _{y})\,\left\{ \left[ d_{xx}(q_{x})-e_{x}\right]
d_{yy}(q_{x})-d_{xy}^{2}(q_{x})\right\} ,
\end{equation}%
and 
\begin{equation}
c_{v}={\frac{a_{x}a_{y}l_{B}^{2}}{(2\pi )^{2}}}\int d{q_{x}}{\frac{%
d_{xx}-e_{x}}{\sqrt{g_{x}(q_{x})}}}.  \label{cv}
\end{equation}%
The quantity $A_{y}$ bears a singular $1/f$ term which is responsible
for the unusual feature of $\tilde{\zeta}_{y}^{\mathrm{ret}}(\omega )$. In
contrast, $A_{x}(f)$ can be easily shown to converge to a constant at small $%
f$. Therefore, in Eq.~(\ref{SPEPRSB}), the large $t$ behavior of $%
I(t,K_{y0})I(t,K_{x0})$ is dominated by $I(t,K_{y0})$, and the two terms
within the braces have qualitatively the same small-$\omega $ behavior.
Substituting Eq.~(\ref{Ay}) in Eq.~(\ref{Ialpha}) leads to 
\begin{equation}
I(t,K_{y0})\sim (1+it\Lambda _{\omega })^{\gamma +2},  \label{Iy1}
\end{equation}%
where 
\begin{equation}
\gamma ={\frac{K_{y0}^{2}c_{v}}{\pi }}-2.  \label{gamma}
\end{equation}%
Here $\Lambda _{\omega }$ is a high-energy cutoff of order the magnetophonon
band width. Inserting Eq.~(\ref{Iy1}) into Eq.~(\ref{SPEPRSB}) and keeping
only the leading-order terms in $\omega $ we produce Eqs.~(\ref%
{rezetasmallwdepin}-\ref{imzetasmallwdepin}) provided Eq. (\ref{inequality})
is met. For larger values of $\gamma $, the solution is not self-consistent,
and one must revert to the full RSB (pinned) solution.

Equations (\ref{rezetasmallwdepin}-\ref{inequality}) are the criteria for the
existence of a PRSB solution. The inequality (\ref{inequality}) defines a
critical value $\gamma_c=1$. Since $\gamma$ in Eq.~(\ref{gamma}) increases
monotonically with $\Delta\nu$, this critical value corresponds to a
critical filling $\Delta\nu_c$. Our numerical result shown below in
Sec.~VI~C indicates that $\Delta\nu_c$ obtained this way matches nicely with
the critical filling defined in the RSB state through the collapse of the
pinning peaks. One can also see that in the vanishing disorder limit ($e_x,
\alpha_y\rightarrow 0$), $\gamma$ reduces to $\gamma_0$ defined in Eq.~(\ref%
{gammaRG}) that occurs in the RG flow equation (\ref{RGflow}), and the
condition (\ref{inequality}) matches the RG condition for the irrelevance of
the disorder. Technically, the reason these coincide originates from the
similar ways in which the Green's function enters in the SPE 
(\ref{SPEPRSB}) and in the calculation of the scaling dimension of the impurity
term in the RG analysis. The minor difference is that the GVM includes the
renormalization of the Green's function by disorder, while the RG analysis,
being perturbative, uses the Green's function for the pure system. 
 
\begin{figure}[h]
\begin{picture}(250,220)
\leavevmode\centering\includegraphics{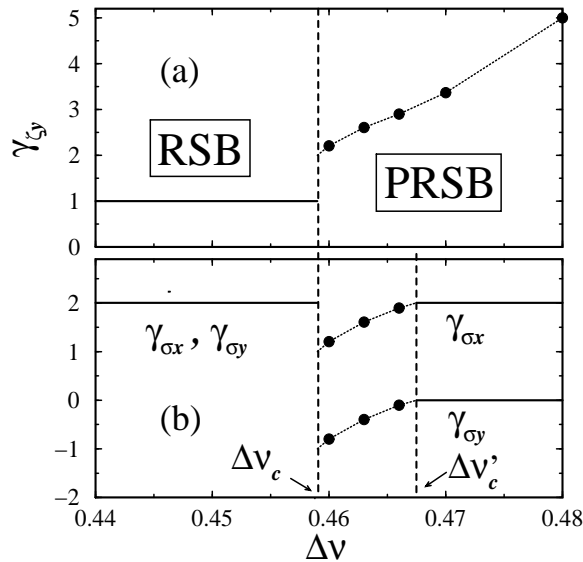}
\end{picture}
\caption{Low-frequency exponents (a) $\protect\gamma _{\protect\zeta y}$ of
Im$\left[ \protect\zeta _{y}^{\mathrm{ret}}\right] $, and (b) $\protect%
\gamma _{\protect\sigma x}$ of Re$\left[ \protect\sigma _{xx}\left( \protect%
\omega \right) \right] $ and $\protect\gamma _{\protect\sigma y}$ of Re$%
\left[ \protect\sigma _{yy}\left( \protect\omega \right) \right] $, as
functions of the partial filling $\Delta \protect\nu $. $v_{\mathrm{imp}}$
is the same as in Fig.~\protect\ref{figsgmxpin}. $\Delta \protect\nu %
_{c}\simeq 0.459$ marks the quantum depinning transition point and $\Delta 
\protect\nu'_{c}$ corresponds to the second transition point at which the
system changes its behavior from divergent as $\protect\omega \rightarrow 0$
to metallic. Circles are the numerical result.}
\label{figexponents}
\end{figure}

In both the RSB and the PRSB states, Im$\left[ \tilde{\zeta}_{y}^{\mathrm{ret%
}}(\omega )\right] $ shows power-law behavior Im$\left[ \tilde{\zeta}_{y}^{%
\mathrm{ret}}(\omega )\right] \sim \omega ^{\gamma _{\zeta y}}$ [Eqs.~(\ref%
{imzetasmallwpin}) and (\ref{imzetasmallwdepin})], although the exponent is
fixed at $1$ for the RSB state. Plotting $\gamma _{\zeta y}$ as a function
of the partial filling in Fig.~\ref{figexponents}~(a), we see that $\gamma
_{\zeta y}$ \emph{jumps} from $1$ in the RSB state to $2$ in the PRSB state.
This jump arises from an underlying jump in the low-frequency exponent in Re$%
\left[ \tilde{\zeta}_{y}^{\mathrm{ret}}(\omega )\right] $ (from $0$ in the
RSB state [Eq.~(\ref{rezetasmallwpin})] to $2$ in the PRSB state [Eq.~(\ref%
{rezetasmallwdepin})]). Such jumps are typical for a KT-type phase
transition. In the next subsection we will show the
corresponding jumps in the low-frequency exponents in conductivities.

\subsection{Anomalous low-frequency exponents for conductivities}

The unusual low-frequency exponents of $\tilde{\zeta}_{y}^{\mathrm{ret}%
}(\omega )$ directly affect the low-frequency behavior of the
conductivities. Inserting Eqs.~(\ref{rezetasmallwpin}-\ref{imzetasmallwpin})
and (\ref{rezetasmallwdepin}-\ref{imzetasmallwdepin}) in Eq.~(\ref%
{longitudinalconductivity}) we find that at small $\omega $, 
\begin{eqnarray}
&&\mathrm{Re}\,\left[ \sigma _{yy}(\omega )\right] \simeq e^{2}\left[
s_{y0}\,\delta (\omega )+s_{y1}\omega ^{\gamma -2}+s_{y2}\right] ,
\label{sgmysmallwPRSB} \\
&&\mathrm{Re}\left[ \,\sigma _{xx}(\omega )\right] \simeq e^{2}\left[
s_{x1}\,\omega ^{\gamma }+s_{x2}\,\omega ^{2}\right] ,
\label{sgmxsmallwPRSB}
\end{eqnarray}%
where $s_{y0}=\Delta \nu e_{x}/2(1-e_{x}\alpha _{y})$, $s_{y1}=e_{x}^{2}%
\beta _{y}\Delta \nu /2\pi $, $s_{y2}=\beta _{x}(1+e_{x}\alpha _{y})\Delta
\nu /2\pi $, $s_{x1}=\beta _{y}(1+e_{x}\alpha _{y})\Delta \nu /2\pi $, $%
s_{x2}=\alpha _{y}^{2}\beta _{x}\Delta \nu /2\pi $. The most significant
feature in Re$\left[ \sigma _{yy}(\omega )\right] $ lies in the $\delta $
peak at $\omega =0$ \cite{footnotedeltapeak}. Physically, this means that 
the PRSB phase is a superconducting state, and the system manages to find an 
effective \textit{free} path to slide along the stripes. By contrast, 
Re$\left[ \sigma_{xx}(\omega \rightarrow 0)\right] =0$, implying the system 
is insulating
for motion perpendicular to the stripes. This suggests that the PRSB state
has \textit{infinite} anisotropy in the DC conductivity. This is not
observed in DC transport experiments \cite{stripesexpr}, and we comment in
Sec.~VIII on what is missing from our model that we believes leads to this
discrepancy.

The other two terms in Eq.~(\ref{sgmysmallwPRSB}) and the terms in Eq.~(\ref%
{sgmxsmallwPRSB}) imply an incoherent contribution at $\omega \neq 0$.
Interestingly, these terms compete with each other in determining the
low-frequency exponents of the conductivities, leading to a second
transition. For $1\leq \gamma <2$, the second term in Eq.~(\ref%
{sgmysmallwPRSB}) and the first term in Eq.~(\ref{sgmxsmallwPRSB}) dominate,
so that Re$\left[ \sigma _{yy}(\omega )\right] \sim \omega ^{-(2-\gamma )}$
which diverges as $\omega \rightarrow 0$ \cite{missed}. This is a very
unusual finite frequency response, which arises from the form of the Green's
function in the PRSB state and so appears to be specific to this system just
after the depinning transition. The response perpendicular to the stripes is
insulating but also anomalous, Re$\left[ \sigma _{xx}\left( \omega \right) %
\right] \sim \omega ^{\gamma }$ with $\gamma $ non-integer. Since $\gamma $
increases with the filling $\Delta \nu $, the low-frequency exponents of Re$%
\left[ \sigma _{yy}(\omega )\right] $ and Re$\left[ \sigma _{xx}(\omega )%
\right] $ evolve (continuously) from $\Delta \nu _{c}$ (for which $\gamma =1$%
) to a second critical filling $\Delta \nu'_{c}$ (for which $\gamma =2$).
As $\Delta \nu $ further increases from $\Delta \nu'_{c}$, $\gamma $
becomes larger than $2$, and the third term in Eq.~(\ref{sgmysmallwPRSB})
and the second term in Eq.~(\ref{sgmxsmallwPRSB}) dominate the low frequency
behavior. Consequently, Re$\left[ \sigma _{yy}(\omega )\right] \sim const.$
for small but non-vanishing $\omega$,
which is a standard finite frequency response for a superconductor,
sometimes called \textquotedblleft incoherent metallic
behavior\textquotedblright\ \cite{kohn}. Furthermore, Re$\left[ \sigma
_{xx}\left( \omega \right) \right] \sim \omega ^{2}$ which is similar to the
behavior in the fully pinned state. Thus, at $\Delta \nu'_{c}$, the system
experiences a second transition in which the finite-frequency behavior of
the stripes changes. The conductivities thus have a very unusual
low-frequency behavior for a small window of filling factors, $\Delta \nu
_{c}<\Delta \nu <\Delta \nu'_{c}$. Interestingly, such changes in power-law
behavior above a KT transition is known to occur in other contexts \cite%
{sudbo}. The qualitative result of the low-frequency exponents $\gamma
_{\sigma \alpha }$ of Re$\left[ \sigma _{\alpha \alpha }\left( \omega
\right) \right] $ discussed here can be seen in Fig.~\ref{figexponents}~(b).
The numerical values of $\Delta \nu _{c}$ and $\Delta \nu'_{c}$ for our
calculations will be discussed in the next subsection. As also shown in Fig.~%
\ref{figexponents}~(b), both $\gamma _{\sigma x}$ and $\gamma _{\sigma y}$
jump at the depinning transition point $\Delta \nu _{c}$.

In practice, the visibility of the various terms in Eqs.~(\ref%
{sgmysmallwPRSB}) and (\ref{sgmxsmallwPRSB}) depends on the relative size of
the coefficients of each term, which we discuss in the next subsection.

\begin{figure}[h]
\begin{picture}(250,190)
\leavevmode\centering\includegraphics{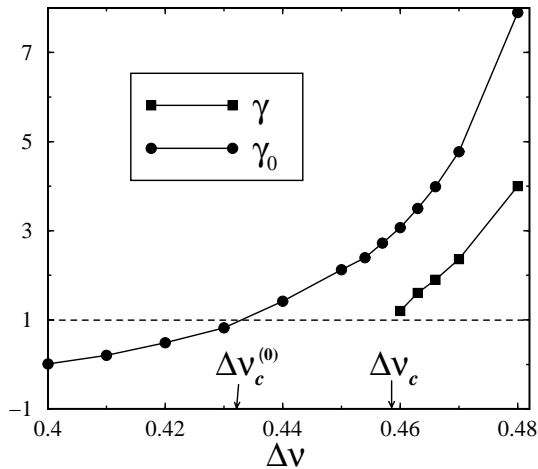}
\end{picture}
\caption{Exponents $\protect\gamma $ and $\protect\gamma _{0}$, defined in
Eqs.~(\protect\ref{gamma}) and (\protect\ref{gammaRG}), respectively, as
functions of $\Delta \protect\nu $. The values $\protect\gamma =1$ and $%
\protect\gamma _{0}=1$ define the depinning transition point $\Delta \protect%
\nu _{c}$ for $v_{imp}=0.0005e^{4}/l_{B}^{2}$ and $\Delta \protect\nu %
_{c}^{(0)}$ in the pure limit, respectively. }
\label{figgamma}
\end{figure}

\subsection{Constraint for $e_x$ and numerical result for the conductivities}

To obtain quantitative results for the conductivities in the PRSB state, we
need to numerically solve the SPEs (Eqs. \ref{SPEs}) together with the
constraint for $e_{x}$. This constraint can be obtained, under the
assumption of the existence of linear-$\omega $ term in $\tilde{\zeta}_{x}^{%
\mathrm{ret}}(\omega )$ at small $\omega $, from the first constraint Eq.~(%
\ref{constraint1}) in the pinned state by examining the limit $%
e_{y}\rightarrow 0$. It is easy to find that $g_{xx}\sim const.$, $%
g_{xy}\sim |e_{y}|^{1/2}$, and $g_{yy}\sim |e_{y}|^{3/2}$. On the other
hand, $e^{-W(\pm K_{x0},0)}$ tends to a non-vanishing constant as $%
e_{y}\rightarrow 0$, while both $e^{-W(0,\pm K_{y0})}$ and 
$e^{-W(\pm K_{x0},\pm K_{y0})}$ scale as $%
|e_{y}|^{-(\gamma +2)/2}$, where $\gamma $ was defined in Eq.~(\ref{gamma}).
Plugging these into Eq.~(\ref{U}) yields $U_{xx}\sim const.$, $U_{xy}\sim
|e_{y}|^{-1/2}$, $U_{yx}\sim |e_{y}|^{(\gamma +1)/2}$, and 
\begin{equation}
U_{yy}\sim |e_{y}|^{(\gamma -1)/2}.  \label{Uyy}
\end{equation}%
Eq.~(\ref{constraint1}) thus becomes 
\begin{equation}
\left( U_{xx}-1\right) \left( c_{0}|e_{y}|^{(\gamma -1)/2}-1\right) \sim
|e_{y}|^{\gamma /2},  \label{constraintPRSB0}
\end{equation}%
where $c_{0}$ is a constant whose precise value is irrelevant for our discussion.
In the PRSB state, $\gamma >1$ and $e_{y}=0$, so that the constraint becomes 
\begin{equation}
U_{xx}-1=0.  \label{constraintPRSB}
\end{equation}%
In Appendix~\ref{solution2}, we discuss another way of satisfying Eq.~(\ref%
{constraintPRSB0}) which leads to an unphysical solution.

\begin{table}[tbp]
\begin{tabular}{|c|c|c|c|c|c|c|c|c|}
\hline
\hspace{0.2cm} $\Delta\nu$ \hspace{0.2cm} & \hspace{0.4cm} $e_x$ \hspace{%
0.4cm} & \hspace{0.2cm} $\beta_x$ \hspace{0.2cm} & \hspace{0.2cm} $\alpha_y$ 
\hspace{0.2cm} & \hspace{0.2cm} $\beta_y$ \hspace{0.2cm} & \hspace{0.2cm} $%
s_{y1}$ \hspace{0.2cm} & \hspace{0.2cm} $s_{y2}$ \hspace{0.2cm} & \hspace{%
0.1cm} $s_{x1}$ \hspace{0.2cm} & \hspace{0.2cm} $s_{x2}$ \hspace{0.2cm} \\ 
\hline
0.46 & -0.0037 & 2.26 & 0.68 & 2.2 & 2.2e-6 & 0.165 & 0.161 & 0.076 \\ \hline
0.463 & -0.0037 & 1.96 & 0.34 & 2.6 & 2.6e-6 & 0.143 & 0.192 & 0.016 \\ 
\hline
0.466 & -0.0037 & 1.98 & 0.18 & 2.9 & 2.9e-6 & 0.146 & 0.215 & 0.0049 \\ 
\hline
0.47 & -0.0035 & 1.97 & 0.10 & 3.4 & 3.1e-6 & 0.147 & 0.252 & 0.0014 \\ 
\hline
0.48 & -0.0034 & 1.85 & 0.02 & 5.0 & 4.4e-6 & 0.141 & 0.382 & 4e-5 \\ \hline
\end{tabular}
\caption{Table of the coefficients of the leading order terms of the self
energy and conductivities in the PRSB state at various filling.}
\end{table}

We have carried out numerical calculations of the exponents and
conductivities for the fillings $\Delta \nu =0.46,0.463,0.466,0.47,0.48$,
all of which are in the PRSB state. Results for $\gamma $ are shown as
squares in Fig.~\ref{figgamma}. Clearly, $\gamma $ increases monotonically
with filling factor. Notice that $\gamma $ at $\Delta \nu =0.46$ is very
close to the critical value $1$, and by an extrapolation we conclude that $%
\Delta \nu _{c}\simeq 0.459$. By comparing with Fig.\ref{figpeak} in the RSB
state, we find that $\Delta \nu _{c}$ agrees with that from the RSB state,
yielding a non-trivial check on our numerics. In Fig.~\ref{figgamma} we also
plot $\gamma _{0}$ (circles) which is computed from Eq.~(\ref{gammaRG}) in
the pure limit. The critical $\gamma _{0}$ results in a critical filling at
the vanishing disorder limit $\Delta \nu _{c}^{(0)}\simeq 0.432$. The result
of $\Delta \nu _{c}>\Delta \nu _{c}^{(0)}$ reflects the fact that a
stronger disorder strength makes pinning more likely and so increases $%
\Delta \nu _{c}$. As we mentioned before, the disorder level we choose is
most likely larger than the experimental situation. We expect that in the
experimental parameter regime the critical filling for the quantum depinning
transition for $N=3$ is some value between $0.432$ and $0.459$.

The numerically computed values of $\gamma_{\zeta y}$, $\gamma_{\sigma x}$
and $\gamma_{\sigma y}$ are shown as circles in Fig.~(\ref{figexponents}).
We find $\Delta\nu'_{c}\simeq 0.467$.

\begin{figure}[h]
\begin{picture}(250,200)
\leavevmode\centering\includegraphics{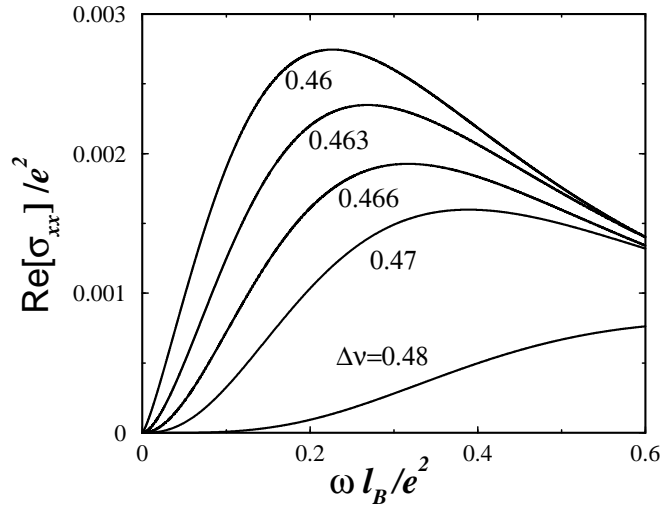}
\end{picture}
\caption{Real part of conductivity perpendicular to the stripes as a
function of frequency in the depinned (PRSB) state. The value $v_{\mathrm{imp%
}}=0.0005e^{4}/l_{B}^{2}$ is used. Curves from left to right correspond to $%
\Delta \protect\nu =0.46,0.463,0.466,0.47$ and $0.48$ respectively. Note
that the low-energy pinning mode is absent in the PRSB.}
\label{sgmxdepin}
\end{figure}

Our numerical result also confirms the low-energy behavior of the
self-energy Eqs.~(\ref{rezetasmallwpin}-\ref{imzetasmallwpin}) and (\ref%
{rezetasmallwdepin}-\ref{imzetasmallwdepin}). In Tab.~1, we present the
coefficients of the leading terms of the self energy. It is clear that $e_x$
is nearly a constant, and $\beta_x$ and $\beta_y$ are also moderate
functions of $\Delta\nu$. But $\alpha_y$ increases drastically as $\Delta\nu$
approaches $\Delta\nu_c$ from above. We will comment on this in Sec.~VII.

It is important to note that the coefficients we find for $s_{y1}$ are
numerically very small (see Table I), so that the anomalous divergence near $%
\omega =0$ can only be visible at very small frequencies. This suggests that
the divergence may be in practice difficult to observe, and indeed it is
beyond the numerical accuracy of our calculations too because the frequency
grid required would be much finer than can practically be achieved. For
frequencies of order $\omega >10^{-5}$ we find that the anomalous divergence
cannot be seen (for the parameters of our calculation) and the finite
frequency response appears to be that of an incoherent metal. Interestingly we
find that the incoherent contribution to the dynamical conductivity becomes
sharply peaked for $\omega <.001$, but this levels off to a constant when
the frequency becomes small enough. 

Numerical results for the conductivities perpendicular to the stripes at
various fillings are shown for a fairly large frequency range in Fig.~\ref%
{sgmxdepin}. The low-energy pinning mode is absent, and there is
instead a broad peak at high frequencies. We would like to remark that the
result at such high energy scales should be taken with a grain of salt, as
our elastic model only reproduces the excitation spectrum of the quantum
Hall stripes for low energies. The peak may be interpreted as being due to a
maximum in the phonon density of states that occurs in the elastic model,
and is \textit{not} a pinning peak. For much lower frequencies where our
computation is accurate, Re$\left[
\sigma _{xx}(\omega )\right] $ shows the power law behavior as expected. For 
$\Delta \nu =0.47$ and $0.48$ which are larger than $\Delta \nu'_{c}$, the
anticipated Re$\left[ \sigma _{xx}(\omega )\right] \simeq \omega ^{2}$ is
not visible within our numerical accuracy. As shown in Table~I, the
coefficient of the $\omega ^{2}$ term, $S_{x2}$, is much smaller than that
of the $\omega ^{\gamma }$ term, $S_{x1}$, for $\Delta \nu =0.47$ and $0.48$%
, again requiring a very fine $\omega $ grid to observe.  Thus in practice one
may observe the anomalous power law dependence over a relatively large
range of filling factors.

\section{Quantum depinning transition - KT universality class}

In the previous section, we observed jumps in the low-frequency exponents of 
$\tilde{\zeta}_{y}^{\mathrm{ret}},$ Re$\left[ \sigma _{xx}\left( \omega
\right) \right] $ and Re$\left[ \sigma _{yy}\left( \omega \right) \right] $
at the quantum depinning transition point. In this section, we discuss the
connection of these jumps with the universal jump in the superfluid
stiffness and the critical exponent of correlation functions of the
KT transition \cite{nelson}.

The depinning transition we have found is of the KT form, as is clear from
the perturbative RG analysis \cite{YFC}. Inserting the smectic form of $%
D_{yy}(\mathbf{q})$ in Eq.~(\ref{smecticDyy}) into the action (\ref{S0eff})
one can see that $d_{yy}(q_x)$ acts as an effective stiffness along the
stripes direction. The action in Eqs.~(\ref{Seff})-(\ref{Simpeff}) for the
stripes phase then can be viewed as a generalized quantum sine-Gordon model:
for $v_{\mathrm{imp}}=0$, the action behaves as a collection of 1+1
dimensional elastic systems, one for each $q_x$; the impurity term couples
these systems. As is well-known, the two-dimensional classical sine-Gordon 
model supports
a roughening transition \cite{chaikin}, which formally is closely related to
a smectic-to-crystal transition, and is a dual description of the KT vortex
unbinding transition \cite{chaikin}.

\begin{figure}[h]
\begin{picture}(250,200)
\leavevmode\centering\includegraphics{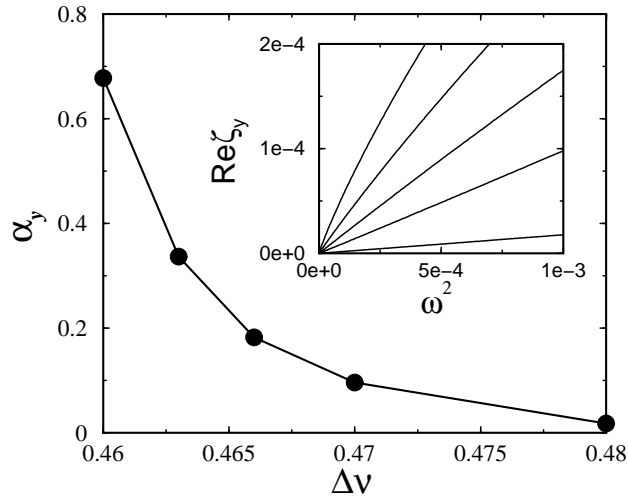}
\end{picture}
\caption{Coefficient of the quadratic in frequency term in Re$\left[ \protect%
\zeta _{y}^{\mathrm{(ret)}}\right] $ as a function of the partial filling in
the depinned (PRSB) state. Inset: Re$\left[ \protect\zeta _{y}^{\mathrm{(ret)%
}}\right] $ as a function of $\protect\omega ^{2}$. Curves from top to
bottom correspond to $\Delta \protect\nu =0.46,0.463,0.466,0.47,0.48$,
respectively. }
\label{rezyPRSB}
\end{figure}

An interesting aspect of our system is that, in the pure limit, there is no
term that is quadratic in $\omega $ in either diagonal component of the
Green's function, so that there is no analogue of $d_{yy}(q_{x})$ in the
time direction. However, such a term \textit{is} generated in the
self-energy as a result of the variational method when the disorder is
present, even in the depinned state. Writing Re$\left[ \tilde{\zeta}_{y}^{%
\mathrm{ret}}(\omega )\right] \simeq \alpha _{y}\omega ^{2}$ for small $%
\omega $, we plot $\alpha _{y}$ as a function of partial filling factor  
$\Delta \nu $ above the transition in Fig.~\ref{rezyPRSB}. One can see the 
sharp increase as the
transition is approached. Such an increase is consistent with the usual RG
for the roughening transition, for which the stiffness increases in the RG
flows, although one needs to go to higher order in perturbation theory than
was undertaken in Ref. \onlinecite{YFC} to see this. We note finally that $%
\alpha _{y}$ \textit{cannot} increase indefinitely: as it increases, the
value of $\gamma $ (Eq. (\ref{gamma})) decreases, eventually crossing the
critical value and forcing the system into the fully pinned state. In this
state, Re$\left[ \tilde{\zeta}_{y}^{\mathrm{ret}}(\omega )\right] \simeq
\alpha _{y}\omega ^{2}$ no longer vanishes as $q_{y},\omega \rightarrow 0$,
but rather goes to a constant. This can be roughly interpreted as a system
with an infinite stiffness, so that one may associate the transition with a
jump in $\alpha _{y}$ from its critical value to infinity.

\section{Conclusion}

In this paper, we have used replicas and the Gaussian variational method to
calculate the finite-frequency conductivity of QH stripes in order to see
its detailed behavior in the vicinity of the quantum depinning transition.
The low-energy degrees of freedom of the QH stripes are described within an
effective elastic model that is characterized by a dynamical matrix which is
determined by matching to microscopic TDHFA calculation. Our results show
that in the pinned state for $\Delta \nu <\Delta \nu _{c}$, the system is in
an RSB state, and the conductivities have resonant peaks for excitation both
parallel and perpendicular to the stripes. As $\Delta \nu $ approaches $%
\Delta \nu _{c}$ from below, a Debye-Waller factor $W(K_{x};K_{y}\neq 0)$
increases and eventually diverges at $\Delta \nu =\Delta \nu _{c}$,
resulting in a vanishing pinning energy $e_{y}$ for motion along the
stripes. For $\Delta \nu >\Delta \nu _{c}$, the system enters a new state
with \textit{partial} replica symmetry breaking (PRSB), in which the
solution has RSB perpendicular to the stripes, but is replica symmetric
along them. In this state Re$\left[ \sigma _{yy}(\omega )\right] $ has a
superconducting response at zero frequency and an anomalous power law
behavior for both Re$\left[ \sigma _{xx}(\omega )\right] $ and Re$\left[
\sigma _{yy}(\omega )\right] $ for $\Delta \nu $ just above the critical
value. Moreover, there are jumps in the low-frequency exponents of both the
self-energy and conductivities at the transition point, as one might expect
for a KT transition.

We conclude by discussing a prominent discrepancy between our results and
those of existing experiments. In DC transport, one observes metallic
behavior with {finite} anisotropy rather than the infinite one found in the
PRSB state. We believe the missing ingredients from our model are processes
allowing hopping of electrons between stripes. These processes are very
difficult to incorporate into an elastic model. It is clear that, if
relevant in the RG sense, such processes can broaden the $\delta $-function
response to yield anisotropic metallic behavior. Our results should apply at
frequency scales above this broadening. Indeed, microwave absorption
experiments become quite challenging at low frequencies, and it is unclear
whether existing measurements of the dynamical conductivity can access the
low frequency conductivity in the unpinned state, whether or not it is
broadened. In any case, it is interesting to speculate that a true $\delta $%
-function response might be accessible in structured environments where
barriers between stripes may suppress electron hopping among stripes \cite%
{Endo}, or that there may be analogous states for layered 2+1 dimensional
classical systems of long string-like objects, which has been shown \cite%
{fertig99} to be closely related to the two-dimensional quantum stripe
problem.

\begin{acknowledgments}
The authors are especially grateful to R.~Lewis, L.~Engel, and Y. Chen for
many stimulating discussions about this problem, and for showing us their
experimental data prior to publication. We are also indebted to
T.~Giamarchi, G.~Murthy, E.~Orignac, E.~Poisson, and A.H.~MacDonald for
useful discussions and suggestions. This work was supported by a NSF Grant
No. DMR-0414290, by a grant from the Fonds Qu\'{e}b\'{e}cois de la recherche
sur la nature et les technologies and a grant from the Natural Sciences and
Engineering Research Council of Canada, and by a grant from SKORE-A program.
\end{acknowledgments}

\appendix

\section{Summary of HF and TDHF formalisms}

\label{HFappendix}

In a previous work\cite{CF}, two of us have shown that, in the HFA, the 
smectic state (as in the edge state model\cite{edgestatemodel}) is unstable 
with respect to density modulations along the
direction of the stripes. The ground state of the two-dimensional electron 
gas near half filling of the higher Landau levels is instead an anisotropic 
two-dimensional Wigner crystal with basis vectors 
$\mathbf{R}_{1}=\left( 0,a_{y}\right) $
and $\mathbf{R}_{2}=\left( a_{x},a_{y}/2\right) $. (One can also see this crystal
as an array of 1D Wigner crystals with out-of-phase modulations on adjacent
1D crystals). The electronic density of this crystal is fully determined by
the Fourier components of the electronic density $\left\{ \left\langle
n\left( \mathbf{K}\right) \right\rangle \right\} $ where $\mathbf{K}$ is a
reciprocal lattice vector of the oblique lattice shown in 
Fig.~\ref{stripeselastic}. 

In our analysis, the Hilbert space is restricted to that of the partially
filled Landau level. It is then convenient to define a density of orbit
centers or \textquotedblleft guiding-center density\textquotedblright\ $%
\left\langle \rho \left( \mathbf{K}\right) \right\rangle $ which is related
to the electronic density by the equation%
\begin{equation}
\left\langle n\left( \mathbf{K}\right) \right\rangle =N_{\phi }F_{N}\left( 
\mathbf{K}\right) \left\langle \rho \left( \mathbf{K}\right) \right\rangle ,
\end{equation}%
where $N_{\varphi }$ is the Landau-level degeneracy and 
\begin{equation}
F_{N}\left( \mathbf{K}\right) =e^{-K^{2}l_{B}^{2}/4}L_{N}^{0}\left( \frac{%
K^{2}l_{B}^{2}}{2}\right) ,
\end{equation}%
( $L_{N}^{0}\left( x\right) $ is a generalized Laguerre polynomial) is a
form factor for an electron in Landau level $N.$ The $\left\langle \rho
\left( \mathbf{K}\right) \right\rangle ^{\prime }s$ can be computed\cite{CF}
by solving the HF\ equation of motion for the single particle Green's
function%
\begin{equation}
{\cal G}\left( \mathbf{K,}\tau \right) =-\frac{1}{N_{\phi }}\sum_{X,X^{\prime }}e^{-%
\frac{i}{2}K_{x}\left( X+X^{\prime }\right) }\delta _{X,X^{\prime
}-K_{y}l_{B}^{2}}\left\langle {\mathcal{T}}_{\tau }c_{X}\left( \tau \right)
c_{X^{\prime }}^{\dagger }\left( 0\right) \right\rangle ,  \label{1_3}
\end{equation}%
with 
\begin{equation}
\left\langle \rho \left( \mathbf{K}\right) \right\rangle ={\cal G}
\left( \mathbf{K,} \tau =0^{-}\right) .
\end{equation}%
In Eq. (\ref{1_3}), $c_{X}\left( c_{X}^{\dagger }\right) $ is the
destruction(creation) operator for an electron in Landau level $N$ with
guiding-center $X$ in the Landau gauge.

From the set of $\left\langle \rho \left( \mathbf{K}\right) \right\rangle
^{\prime }s$ computed in the HFA, one can derive the dynamical
\textquotedblleft density-density\textquotedblright\ correlation function%
\begin{equation}
\chi _{\mathbf{K},\mathbf{K}^{\prime }}^{\left( \rho ,\rho \right) }\left( 
\mathbf{q},\tau \right) =-N_{\varphi }\left\langle {\mathcal{T}}_{\tau }%
\widetilde{\rho }\left( \mathbf{q}+\mathbf{K},\tau \right) \widetilde{\rho }%
\left( -\mathbf{q}-\mathbf{K}^{\prime },0\right) \right\rangle ,  \label{1_4}
\end{equation}%
in the TDHFA \cite{CF}. In Eq.~(\ref{1_4}), $\mathbf{q}$ is a vector restricted
to the first Brillouin zone
of the stripe crystal and $\widetilde{\rho }\equiv \rho -\left\langle \rho
\right\rangle $. By following the poles of $\chi^{\left( \rho ,\rho \right) }$
with non-vanishing weight as the
wavevector $\mathbf{q}$ is varied in the Brillouin zone of the reciprocal
lattice, we get the dispersion relation of the phonon and higher-energy
collective modes of the stripe state. The equation of motion for $\chi _{%
\mathbf{K},\mathbf{K}^{\prime }}^{\left( \rho ,\rho \right) }\left( \mathbf{q%
},\tau \right) $, in the TDHFA, is given by 
\begin{equation}
\sum_{\mathbf{K}^{\prime \prime }}\left[ i\omega _{n}\delta _{\mathbf{K},%
\mathbf{K}^{\prime }}-M_{\mathbf{K},\mathbf{K}^{\prime \prime }}\left( 
\mathbf{q}\right) \right] \chi _{\mathbf{K}^{\prime \prime },\mathbf{K}%
^{\prime }}^{\left( \rho ,\rho \right) }\left( \mathbf{q},i\omega
_{n}\right) =B_{\mathbf{K},\mathbf{K}^{\prime }}\left( \mathbf{q}\right) ,
\label{7p1}
\end{equation}%
where $\omega _{n}$ is a Matsubara bosonic frequency and the matrices $M_{%
\mathbf{K},\mathbf{K}^{\prime }}$ and $B_{\mathbf{K},\mathbf{K}^{\prime }}$
are defined by%
\begin{eqnarray}
M_{\mathbf{K},\mathbf{K}^{\prime }}\left( \mathbf{q}\right) &=&-2i\left( 
\frac{e^{2}}{\kappa l_{B}}\right) \left\langle \rho \left( \mathbf{K-K}%
^{\prime }\right) \right\rangle  \label{rene1} \\
&&\times \sin \left[ \frac{\left( \mathbf{q}+\mathbf{K}\right) \times \left( 
\mathbf{q}+\mathbf{K}^{\prime }\right) l_{B}^{2}}{2}\right] \left[
H_{N}\left( \mathbf{K}-\mathbf{K}^{\prime }\right) -X_{N}\left( \mathbf{K-K}%
^{\prime }\right) -H_{N}\left( \mathbf{q}+\mathbf{K}^{\prime }\right)
+X_{N}\left( \mathbf{q}+\mathbf{K}^{\prime }\right) \right]  \notag
\end{eqnarray}%
\newline
and%
\begin{equation}
B_{\mathbf{K},\mathbf{K}^{\prime }}\left( \mathbf{k}\right) =2i\sin \left[ 
\frac{\left( \mathbf{q}+\mathbf{K}\right) \times \left( \mathbf{q}+\mathbf{K}%
^{\prime }\right) l_{B}^{2}}{2}\right] \left\langle \rho \left( \mathbf{K-K}%
^{\prime }\right) \right\rangle
\end{equation}%
respectively. (Here $\mathbf{a}\times \mathbf{b}$ stands for $a_xb_y-a_yb_x$.)

In Eq. (\ref{rene1}), $H_{N}\left( \mathbf{q}\right) $ and $X_{N}\left( 
\mathbf{q}\right) $ are the HF interactions in Landau level $N$%
:%
\begin{eqnarray}
H_{N}\left( \mathbf{q}\right) &=&\left( \frac{e^{2}}{\kappa l_{B}}\right) 
\frac{1}{ql_{B}}e^{\frac{-q^{2}l_{B}^{2}}{2}}\left[ L_{N}^{0}\left( \frac{%
q^{2}l_{B}^{2}}{2}\right) \right] ^{2}, \\
X_{N}\left( \mathbf{q}\right) &=&\left( \frac{e^{2}}{\kappa l_{B}}\right) 
\sqrt{2}\int_{0}^{\infty }dx\,e^{-x^{2}}\left[ L_{N}^{0}\left( x^{2}\right) %
\right] ^{2}J_{0}\left( \sqrt{2}xql_{B}\right) .
\end{eqnarray}

To solve for $\chi _{\mathbf{K},\mathbf{K}^{\prime }}^{\left( \rho ,\rho
\right) }\left( \mathbf{q},i\omega _{n}\right) $, we diagonalize the matrix $%
M_{\mathbf{K},\mathbf{K}^{\prime \prime }}\left( \mathbf{q}\right) $ by the
transformation

\begin{equation}
M=CEC^{-1},  \label{eqnm}
\end{equation}%
where $C$ is the matrix of the eigenvectors of $M$ and $E_{i,j}=\varepsilon
_{j}\delta _{i,j}$ is the diagonal matrix of its eigenvalues. The analytic
continuation of $\chi _{\mathbf{K},\mathbf{K}^{\prime }}^{\left( \rho ,\rho
\right) }\left( \mathbf{q},i\omega _{n}\right) $ is given by 
\begin{eqnarray}
\chi _{\mathbf{K},\mathbf{K}^{\prime }}^{\left( \rho ,\rho \right) }\left( 
\mathbf{q},\omega \right) &=&\sum_{j,k}\frac{C_{\mathbf{K},j}\left( \mathbf{q%
}\right) \left[ C\left( \mathbf{q}\right) ^{-1}\right] _{j,k}B_{k,\mathbf{K}%
^{\prime }}\left( \mathbf{q}\right) }{\omega +i\delta -\varepsilon
_{j}\left( \mathbf{q}\right) }  \label{1_12} \\
&\equiv &\sum_{i}\frac{W_{i}\left( \mathbf{q}+\mathbf{K},\mathbf{q}+\mathbf{K%
}^{\prime }\right) }{\omega +i\delta -\varepsilon _{i}},
\end{eqnarray}%
where $W_{i}\left( \mathbf{q}+\mathbf{K},\mathbf{q}+\mathbf{K}^{\prime
}\right) $ is the weight of the pole $\varepsilon _{i}$ in the response
function. The true density response function is simply 
\begin{equation}
\chi _{\mathbf{K},\mathbf{K}^{\prime }}^{\left( n,n\right) }\left( \mathbf{q}%
,\omega \right) =N_{\phi }\sum_{i}\frac{F_{N}\left( \mathbf{q}+\mathbf{K}%
\right) W_{i}\left( \mathbf{q}+\mathbf{K},\mathbf{q}+\mathbf{K}^{\prime
}\right) F_{N}\left( \mathbf{q}+\mathbf{K}^{\prime }\right) }{\omega
+i\delta -\varepsilon _{i}}.  \label{21_13}
\end{equation}

\section{Inversion rules for matrices}

\label{inversionrules} 

The inversion rules for hierarchical matrices in the $n \rightarrow 0$ limit
for the case where the entries are scalars may be found in Ref.~\cite{MP}.
In this appendix we generalize these inversion rules for the situation when
the entries are themselves $n_0 \times n_0$ matrices, with our problem
corresponding to $n_0=2$.

\begin{figure}[h]
\begin{picture}(200,160)
\leavevmode\centering\includegraphics{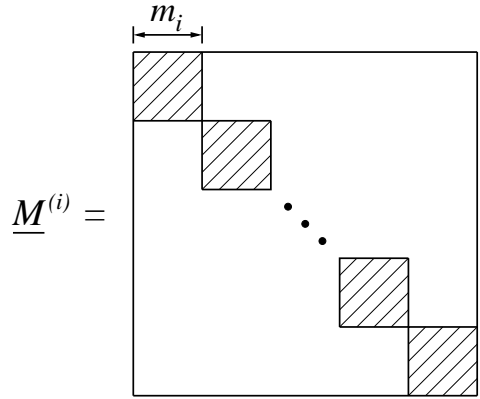}
\end{picture}
\caption{Schematic structure of the matrix $\underline{M}^{(i)}$. All the matrix
elements in the shaded area are $1$ and 0 elsewhere.}
\label{matrix}
\end{figure}

In the replica method, we introduce $n$ replicas of the system and thus deal
with$\left( n_{0}n\right) \times \left( n_{0}n\right) $ matrices. In the RSB
states, in order to invert matrices in the limit of $n\rightarrow 0$
analytically, we follow the scalar case to assume that the $\left(
n_{0}n\right) \times \left( n_{0}n\right) $ matrices have a hierarchical
structure. This may be described by a set of integers $m_{0}(=n),m_{1},%
\cdots ,m_{k},m_{k+1}(=1)$ where $m_{i}/m_{i+1}$ is also an integer. Such
matrices may be constructed by introducing $k+2$ \textquotedblleft
block\textquotedblright\ matrices $\underline{M}^{(i)}$ ($i=0,\cdots ,k+1$)
all of size $n\times n$. These are defined such that their elements
are $1$ within the $m_{i}$ blocks along the diagonal and $0$ elsewhere. 
These matrices can be used
as a basis for a group, so that any $n_{0}n$ by $n_{0}n$ hierarchical matrix 
$\underline{\underline{A}}$ can be expressed as 
\begin{equation}
\underline{\underline{A}}=\tilde{\hat{a}}\otimes \underline{1}+\sum_{i=0}^{k}%
\hat{a}_{i}\otimes \lbrack \underline{M}^{(i)}-\underline{M}^{(i+1)}].
\end{equation}%
where $\tilde{\hat{a}}$ and $\hat{a}_{i}$ are $n_{0}\times n_{0}$ matrices.
It is easy to check that 
\begin{equation}
(\hat{a}_{i}\otimes \underline{M}^{(j)})\cdot (\hat{a}_{l}\otimes \underline{%
M}^{(s)})=(\hat{a}_{i}\cdot \hat{a}_{l})\otimes (\underline{M}^{(j)}\cdot 
\underline{M}^{(s)}).
\end{equation}%
This means that $\underline{\underline{A}}$ is characterized by $k+2$ $n_{0}$
by $n_{0}$ matrices $\tilde{\hat{a}}$ and $\hat{a_{i}}$ ($i=0,\cdots ,k$).
In fact, $\underline{\underline{A}}$ is completely parametrized by its
topmost row 
\begin{eqnarray}
&&\tilde{\hat{a}}\;\;\;\underbrace{\hat{a_{k}}\,\cdots \,\hat{a_{k}}}\;\;\;%
\underbrace{\hat{a_{k-1}}\,\cdots \,\hat{a_{k-1}}}\;\;\;\cdots \cdots \;\;\;%
\underbrace{\hat{a_{0}}\,\cdots \,\hat{a_{0}}}. \\
&&~~~~~~~~~m_{k}~~~~~~~~~~~~~m_{k-1}~~~~~~~~~~~~~~~~~~~~~m_{0}  \notag
\end{eqnarray}%
We can then define 
\begin{equation}
\hat{a}(u)=\left\{ 
\begin{array}{ll}
\hat{a_{0}}\;\; & \mathrm{for}\;n-m_{1}<u<n \\ 
& \vdots \;\;\;\; \\ 
\hat{a_{k}}\;\; & \mathrm{for}\;1<u<m_{k}%
\end{array}%
\right.
\end{equation}%
to parameterize the off-diagonal element matrices.

We assume that the matrix $\underline{\underline{A}}$ has an inverse matrix $%
\underline{\underline{B}}$ which because of the group properties should also
be a hierarchical matrix, and thus is characterized by $\tilde{\hat{b}}$ and 
$\hat{b_{i}}$ ($i=1,\cdots ,k$). If we multiply two matrices $\underline{%
\underline{A}}$ and $\underline{\underline{B}}$ and call the product $%
\underline{\underline{C}}$, it may be written as 
\begin{equation}
\underline{\underline{C}}=\underline{\underline{A}}\cdot \underline{%
\underline{B}}=\tilde{\hat{c}}\otimes \underline{1}+\sum_{i=0}^{k}\hat{c_{i}}%
\otimes \lbrack \underline{M}^{(i)}-\underline{M}^{(i+1)}],
\end{equation}%
with 
\begin{eqnarray}
&&\tilde{\hat{c}}=\tilde{\hat{a}}\cdot \tilde{\hat{b}}%
-\sum_{i=0}^{k}(m_{i+1}-m_{i})\hat{a_{i}}\cdot \hat{b_{i}},  \label{tildec0}
\\
&&\hat{c_{i}}=\hat{a_{i}}\cdot \tilde{\hat{b}}-m_{i}\hat{a_{i}}\cdot \hat{%
b_{i}}+\sum_{j=i+1}^{k}(m_{j}-m_{j+1})(\hat{a_{i}}\cdot \hat{b_{j}}+\hat{%
a_{j}}\cdot \hat{b_{i}})-\sum_{j=0}^{i}(m_{j+1}-m_{j})\hat{a_{j}}\cdot \hat{%
b_{j}}.  \label{ci}
\end{eqnarray}

Now we are in the position to analytically continue the hierarchical matrix
to $n\rightarrow 0$. We first analytically continue $\hat{a}(u)$ to be
defined for $u\in \lbrack 1,n]$ and then take the limit $n\rightarrow 0$.
The limit $n\rightarrow 0$ then suggests that the hierarchical matrix $%
\underline{\underline{A}}$ is specified by a diagonal-element matrix $\tilde{%
\hat{a}}$ and a matrix function $\hat{a}(u)$ for $u\in \lbrack 0,1]$. The
matrix $\underline{\underline{B}}$ can be analytically continued in the same
way. Eqs.~(\ref{tildec0}-\ref{ci}) therefore become 
\begin{eqnarray}
&&\tilde{\hat{c}}=\tilde{\hat{a}}\cdot \tilde{\hat{b}}-\int_{0}^{1}du\,\hat{a%
}(u)\cdot \hat{b}(u),  \label{tildec} \\
&&\hat{c}(u)=(\tilde{\hat{a}}-\left\langle \hat{a}\right\rangle )\cdot \hat{b%
}(u)+\hat{a}(u)\cdot (\tilde{\hat{b}}-\langle \hat{b}\rangle )-\int_{0}^{u}dv%
\left[ \hat{a}(u)-\hat{a}(v)\right] \cdot \left[ \hat{b}(u)-\hat{b}(v)\right]
,  \label{cu}
\end{eqnarray}%
where $\left\langle \hat{a}\right\rangle =\int_{0}^{1}dv\,\hat{a}(v)$. Since 
$\underline{\underline{B}}$ is the inverse matrix of $\underline{\underline{A%
}}$, we require 
\begin{equation}
\tilde{\hat{c}}=\hat{1},\;\;\;\;\hat{c}(u)=\hat{0}.  \label{inverse}
\end{equation}%
Differentiating Eq.~(\ref{cu}) with respect to $u$ and using Eq.~(\ref%
{inverse}) leads to 
\begin{equation}
\left\{ \tilde{\hat{a}}-\left\langle \hat{a}\right\rangle -\left[ \hat{a}%
\right] (u)\right\} \cdot \hat{b}^{\prime }(u)+\hat{a}^{\prime }(u)\cdot
\left\{ \tilde{\hat{b}}-\langle \hat{b}\rangle -[\hat{b}](u)\right\} =\hat{0}%
,  \label{p1}
\end{equation}%
where $\left[ \hat{a}\right] (u)=\int_{0}^{u}dv\left[ \hat{a}(u)-\hat{a}(v)%
\right] $, and $\hat{a}^{\prime }(u)=d\hat{a}(u)/du$. By making use of $([%
\hat{a}](u))^{\prime }=u\hat{a}^{\prime }(u)$, Eq.~(\ref{p1}) becomes 
\begin{equation}
\left\{ \tilde{\hat{a}}-\left\langle \hat{a}\right\rangle -\left[ \hat{a}%
\right] (u)\right\} \cdot \left\{ \tilde{\hat{b}}-\langle \hat{b}\rangle -[%
\hat{b}](u)\right\} =const.  \label{p2}
\end{equation}%
To determine the constant matrix in Eq.~(\ref{p2}), we examine Eqs.~(\ref%
{tildec}) and (\ref{cu}) at $u=1$ and get 
\begin{equation}
\left[ \tilde{\hat{a}}-\hat{a}(1)\right] \cdot \left[ \tilde{\hat{b}}-\hat{b}%
(1)\right] =\hat{1}.  \label{p3}
\end{equation}%
So $const.=\hat{1}$, and Eq.~(\ref{p2}) gives 
\begin{equation}
\left\{ \tilde{\hat{b}}-\langle \hat{b}\rangle -[\hat{b}](u)\right\}
=\left\{ \tilde{\hat{a}}-\langle \hat{a}\rangle -[\hat{a}](u)\right\} ^{-1},
\end{equation}%
which can be inserted into Eq.~(\ref{p1}) to produce one of the inversion
rules 
\begin{equation}
\hat{b}(u)-\hat{b}(v)=\int_{u}^{v}dy\,\left\{ \tilde{\hat{a}}-\langle \hat{a}%
\rangle -[\hat{a}](y)\right\} ^{-1}\cdot \hat{a}^{\prime }(y)\cdot \left\{ 
\tilde{\hat{a}}-\langle \hat{a}\rangle -[\hat{a}](y)\right\} ^{-1}.
\label{inversion2}
\end{equation}%
This is very similar to Eq.~(AII.5) in Ref.~ \onlinecite{MP}. Eqs.~(\ref%
{inversion2}) and (\ref{p3}) lead to 
\begin{equation}
\tilde{\hat{b}}-\hat{b}(u)=\left[ \tilde{\hat{a}}-\hat{a}(1)\right]
^{-1}-\int_{u}^{1}dv\,\left\{ \tilde{\hat{a}}-\langle \hat{a}\rangle -[\hat{a%
}](v)\right\} ^{-1}\cdot \hat{a}^{\prime }(v)\cdot \left\{ \tilde{\hat{a}}%
-\langle \hat{a}\rangle -[\hat{a}](v)\right\} ^{-1}.  \label{inversion3}
\end{equation}%
This is the inversion rule we have used in our work [see Eq.~(\ref%
{Ginversion}) in the text].

For completeness, we also show, without giving the details of the
derivation, some other inversion rules as well as the formula for $%
\lim_{n\rightarrow 0}\left( {\frac{1}{n}}\mathrm{Tr}\,\mathrm{ln}\,%
\underline{\underline{A}}\right) $ which appears in the expression of free
energy: 
\begin{eqnarray}
&&\tilde{\hat{b}}=\left( \tilde{\hat{a}}-\langle \hat{a}\rangle \right)
^{-1}\cdot \left\{ \hat{1}-\int_{0}^{1}{\frac{du}{u^{2}}}\,[\hat{a}](u)\cdot
\left( \tilde{\hat{a}}-\langle \hat{a}\rangle -[\hat{a}](u)\right) ^{-1}-%
\hat{a}(0)\cdot \left( \tilde{\hat{a}}-\langle \hat{a}\rangle \right)
^{-1}\right\} ,  \label{inversion1} \\
&&\hat{b}(u)=-\left( \tilde{\hat{a}}-\langle \hat{a}\rangle \right)
^{-1}\cdot \left\{ \hat{a}(0)\cdot \left( \tilde{\hat{a}}-\langle \hat{a}%
\rangle \right) ^{-1}+{\frac{1}{u}}\,[\hat{a}](u)\cdot \left( \tilde{\hat{a}}%
-\langle \hat{a}\rangle -[\hat{a}](u)\right) ^{-1}\right\}  \notag \\
&&\lim_{n\rightarrow 0}\left( {\frac{1}{n}}\mathrm{Tr}\,\mathrm{ln}\,%
\underline{\underline{A}}\right) =\mathrm{ln}\,\mathrm{det}\,\left( \tilde{%
\hat{a}}-\langle \hat{a}\rangle \right) +\mathrm{Tr}\,\left[ \hat{a}(0)\cdot
\left( \tilde{\hat{a}}-\langle \hat{a}\rangle \right) ^{-1}\right]
-\int_{0}^{1}{\frac{du}{u^{2}}}\;\mathrm{ln}\left[ {\frac{\mathrm{det}\left( 
\tilde{\hat{a}}-\langle \hat{a}\rangle -[\hat{a}](u)\right) }{\mathrm{det}%
\left( \tilde{\hat{a}}-\langle \hat{a}\rangle \right) }}\right] .
\end{eqnarray}

\begin{figure}[h]
\begin{picture}(250,320)
\leavevmode\centering\includegraphics{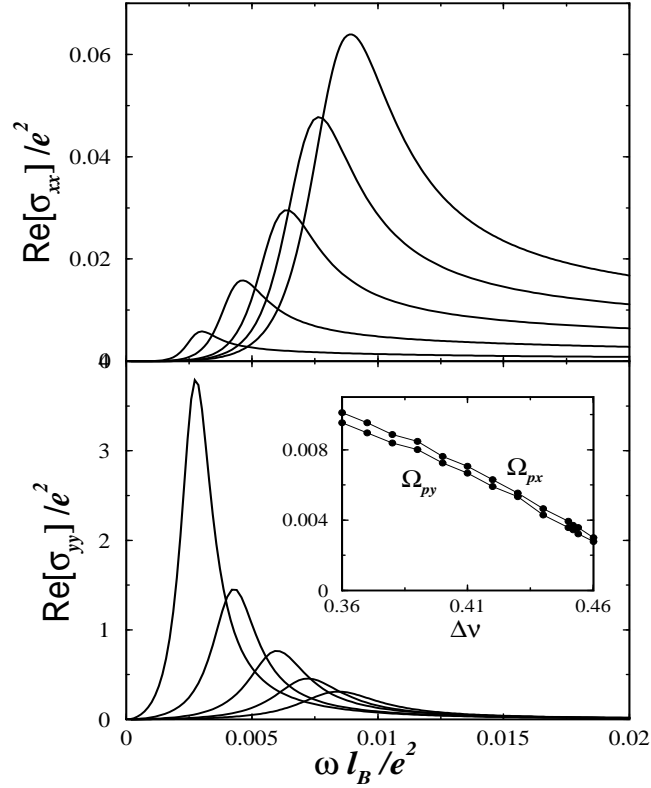}
\end{picture}
\caption{Real part of conductivity perpendicular to the stripes (a) and
parallel to the stripes (b), in the unphysical solution. Curves from right
to left correspond to $\Delta \protect\nu =0.46,0.44,0.42,0.4,0.38$,
respectively. Smaller disorder strength $v_{\mathrm{imp}%
}=0.0001e^{4}/l_{B}^{2}$ is used. Inset in (b): Peak frequencies as
functions of the partial filling.}
\label{unphyssolution}
\end{figure}

\section{SPE's for the retarded self energy}

\label{Appendixanalycon}

In this Appendix, we analytically continue the SPE's (\ref{zetatilde}) for
the Matsubara self energy in order to derive the SPE's (\ref{SPEs}) for the
retarded self energy. We rewrite Eq.~(\ref{zetatilde}) as 
\begin{equation}
\tilde{\zeta}_{\alpha }(\omega _{n}) = 
\int_{0}^{1}du\,\zeta _{\alpha}(u)+4v_{\mathrm{imp}}\int_{0}^{1/T}
d\tau V_{\alpha \alpha }^{\prime }\left[ \widetilde{B}(\tau )\right]
- 4 v_{\mathrm{imp}} J(\omega _{n}),
\label{zetaJ}
\end{equation}%
where 
$J(\omega _{n})$ is the Fourier transform in Matsubara frequencies of 
\begin{equation}
J_{0}(\tau )=\exp \left[-{1\over 2}\sum_{\mu }K_{\mu }^{2}
\widetilde{B}_{\mu \mu }(\tau ) \right]  \label{Jtau}
\end{equation}%
with $\widetilde{B}_{\mu \mu }(\tau )$ being defined in Eq.~(\ref{Btilde}).
Obviously, $J_{0}(\tau )$ is a Matsubara correlation function, and
its corresponding real-time ordered correlation function reads 
\begin{equation}
\tilde{J}_{0}(t) = i J_{0}(\tau \rightarrow it) ={\cal \theta}(t) J_{1}(t)
+ {\cal \theta}(-t) J_{2}(t), \label{Jt}
\end{equation}
where $J_{1}(t)=\tilde{J}_{0}(t>0)$, $J_{2}(t)=\tilde{J}_{0}(t<0)$
with the relation $J_{2}(-t)/i = (J_{2}(t)/i)^*$,
and ${\cal \theta}(t)$ is the step function.
The retarded function becomes
\begin{equation}
J^{\rm ret}_{0}(t) = {\cal \theta}(t) [J_{1}(t) - J_{2}(t)].  \label{Jret}
\end{equation}

Using the Nambu representation 
\begin{equation}
\widetilde{G}_{\mu \nu }(\mathbf{q},\omega _{n})=-{\frac{1}{\pi }}%
\int_{-\infty }^{\infty }df \, {\frac{A_\mu(f)}{i\omega _{n}-f}} ,
\end{equation}%
where $A_\mu(f)$ is defined in Eq.~(\ref{spectralfunction}), 
we find that $\widetilde{B}_{\mu \mu }(\tau )$ in Eq.~(\ref{Btilde}) becomes
\begin{equation}
\widetilde{B}_{\mu \mu }(\tau )={\frac{1}{\pi }}\int_{0}^{\infty }dfA_{\mu
}(f)\left[ T \sum_{\omega_n}(1-\cos \omega_n \tau) {2f\over \omega_n^2+f^2}
\right] .
\end{equation}
We can then easily sum over the Matsubara frequency in the above equation
to get
\begin{equation}
\widetilde{B}_{\mu \mu }(\tau )={\frac{1}{\pi }}\int_{0}^{\infty }dfA_{\mu
}(f)\left[ 1-e^{-|\tau |f}+\frac{2\left[ 1-\cosh (f\tau )\right] }{e^{u/T}-1}%
\right] . \label{BB}
\end{equation}%
Apparently, at $T=0$, the last term inside the parentheses in Eq.~(\ref{BB}) 
vanishes. Inserting Eq.~(\ref{BB}) into Eq.~(\ref{Jtau}) and following the 
procedure described in Eqs.~(\ref{Jt}) and (\ref{Jret}), we find that  
 at $T=0$, 
\begin{equation}
J^{\rm ret}_{0}(t) = i \, {\cal \theta}(t) \, {\rm Im} \, 
\exp \left[-\sum_{\mu }{K_{\mu }^{2}\over {\pi }}\int_{0}^{\infty }dfA_{\mu
}(f)\left( 1-e^{-|\tau |f} \right) \right]. 
 \label{Jret1}
\end{equation}
From Eqs.~(\ref{zetaJ}) and (\ref{Jret1}) and noting that 
$J^{\rm ret}_{0}(\omega)= \int^{-\infty}_\infty dt e^{i\omega t} 
J^{\rm ret}_{0}(t)$, we immediately obtain Eq.~(\ref{SPEs}).

\section{Unphysical solution of the SPE's}

\label{solution2}

In our numerical search, we also notice the existence of another solution
which we present here and argue is unphysical. Fig.~\ref{unphyssolution}
shows the result of the conductivities from this solution which corresponds
to a much smaller disorder level. Both Re$\left[ \sigma _{xx}\left( \omega
\right) \right] $ and Re$\left[ \sigma _{yy}\left( \omega \right) \right] $
show the pinning behavior and the peak frequencies move in as also shown in
the inset of Fig.~\ref{unphyssolution}~(b). However, unlike in the solution
we presented in the text, no quantum depinning transition occurs.

Interestingly, this solution displays a peak move-in behavior that is
reminiscent of what is seen in the physical solution. This is the result of a
decreasing $e_y$, due to the increasing $W(K_x;K_y\neq 0)$ with the partial
filling $\Delta\nu$. However, at small $e_y$, unlike in the other solution,
the constraint (\ref{constraintPRSB0}) is not satisfied through $U_{xx}=1$,
but instead through $U_{yy}=1$. This can be seen from Eq.~(\ref{Uyy}),
according to which $U_{yy}$ will rapidly decrease from very large values to
very small values right near $\gamma=1$. This means that near this value $%
U_{yy}$ must pass through one, satisfying the constraint. In this solution, $%
\gamma$ remains very close to one over a range of filling factors, and does
so by making $|e_x|$ very large, even for small $v_{\mathrm{imp}}$. This
implies an unphysically large pinning for sliding perpendicular to the
stripes. Because of this, and the close agreement between the other solution
and the perturbative RG results, we ignore this solution to the SPE's as
physically unreasonable.


\begin{thebibliography}{99}
\bibitem{stripestheory} A.~A.~Koulakov, M.~M.~Fogler, and B.~I.~Shklovskii,
Phys. Rev. Lett. 76, 499 (1996); M.~M.~Fogler, A.~A.~Koulakov, and
B.~I.~Shklovskii, Phys. Rev. B 54, 1853 (1996); R.~Moessner and
J.~T.~Chalker, \emph{ibid.} 54, 5006 (1996).

\bibitem{AG} I. L. Aleiner and L. I. Glazman, Phys.\ Rev.\ B \textbf{52},
11296 (1995).

\bibitem{DMRG} N. Shibata and D. Yoshioka, Phys.\ Rev.\ Lett.\ \textbf{86},
5755 (2001).

\bibitem{exactdiag} E. H. Rezayi, F. D. M. Haldane, and K. Yang, Phys.\
Rev.\ Lett.\ \textbf{83}, 1219 (1999); F. D. M. Haldane, E. H. Rezayi, and
K. Yang, \textit{ibid.}, \textbf{85}, 5396 (2000).

\bibitem{stripesexpr} M.~P. Lilly \emph{et al}, Phys.\ Rev.\ Lett.\ \textbf{%
82}, 394 (1999); \textbf{83}, 820 (1999); R.~R. Du \emph{et al}, Solid State
Commun.\ \textbf{109}, 389 (1999); W. Pan \emph{et al}, Phys.\ Rev.\ Lett.\ 
\textbf{83}, 820 (1999); K.~B. Cooper \emph{et al}, Phys.\ Rev.\ B \textbf{60%
}, 11285 (1999).

\bibitem{fradkin} E. Fradkin and S. A. Kivelson, Phys. Rev. B \textbf{59},
8065 (1999).

\bibitem{CF} R. C\^{o}t\'{e} and H. A. Fertig, Phys. Rev. B 62, 1993 (2000).

\bibitem{edgestatemodel} A.H. MacDonald and M.P. A. Fisher, Phys. Rev. B 
\textbf{61}, 5724 (2000); A. Lopatnikova \emph{et al.}, \emph{ibid.} \textbf{%
64}, 155301 (2001).

\bibitem{voit} For a review, see J. Voit, Rep. Prog. Phys. \textbf{58},
977 (1995).

\bibitem{kolomeisky} E.B. Kolomeisky and Joseph P. Straley, Rev. Mod. Phys. 
\textbf{68}, 175 (1996).

\bibitem{fertig99} H. A. Fertig, Phys. Rev. Lett. \textbf{82}, 3693 (1999).

\bibitem{FL1} H. Fukuyama and P. A. Lee, Phys. Rev. B \textbf{17}, 535
(1978).

\bibitem{Larkin} A. I. Larkin, Sov. Phys. JETP \textbf{31}, 784 (1970); A.
I. Larkin and Y. N. Ovchinnikov, J. Low. Temp. Phys. \textbf{34}, 409 (1979).

\bibitem{KF} T. Giamarchi and H.J. Schulz, Phys. Rev. B \textbf{37}, 325
(1988); C. Kane and M.P.A. Fisher, Phys. Rev. Lett. \textbf{68}, 1220 (1992).

\bibitem{muwave} R.M. Lewis, P.D. Ye, L.W. Engel, D.C. Tsui, L.N. Pfeiffer,
and K.W. West, Phys. Rev. Lett. 89, 136804 (2002) .

\bibitem{Florida} R. Lewis, L. Engel and Y. Chen, private communication.

\bibitem{YFC} Hangmo Yi, H. A. Fertig, and R. C\^{o}t\'{e}, Phys. Rev. Lett.
85, 4156 (2000). Note in that letter, the stripes are assumed to be along
the $\hat{x}$ direction.

\bibitem{LFCY} M.-R. Li, H.A. Fertig, R. C\^{o}t\'{e}, and H. Yi, Phys. Rev.
Lett. \textbf{92}, 186804 (2004).

\bibitem{MPV} M. M\'ezard, G. Parisi, and M. Virasoro, \emph{Spin Glass
Theory and Beyond} (World Scientific, Singapore, 1987).

\bibitem{MP} M. M\'ezard and G. Parisi, J. Phys. I 1, 809 (1991).

\bibitem{GLD96} T. Giamarchi and P. Le Doussal, Phys. Rev. B 53, 15206
(1996).

\bibitem{GLD95} T. Giamarchi and P. Le Doussal, Phys. Rev. B 52, 1242 (1995).
The replica and GVM was applied to vortex lattice system to predict the
Bragg glass phase, which was later on supported by Monte Carlo
simulations [M. J. P. Gingras and D. A. Huse,
Phys. Rev. B {\bf 53}, 15193 (1996).].

\bibitem{GO} T. Giamarchi and E. Orignac, in \emph{Theoretical Methods for
Strongly Correlated Electrons}, CRM Series in Mathematical Physics (Spring,
Berlin) (also cond-mat/0005220).

\bibitem{CGLD} R. Chitra, T. Giamarchi and P. Le Doussal, Phys. Rev. Lett. 
\textbf{80}, 3827 (1998); Phys. Rev. B \textbf{65}, 035312 (2002). 
These works address the pinned Wigner crystal problem by using the  
replica and GVM and reveal a quite broad pinning peak. Whereas 
some other approaches obtain a very narrow pinning peak [H.~A.~Fertig, 
Phys. Rev. B \textbf{59}, 2120 (1999); M.~M.~Fogler and D.~A.~Huse, 
\emph{ibid.} \textbf{62}, 7553 (2000)]. We have
found that the latter approach is consistent with the former two in the weak
disorder limit. The underlying cause of the narrowing is due to a paucity of
low-lying phonon states in the isotropic crystal, which does not occur for
the stripe state\cite{CF}.

\bibitem{OC} E.~Orignac and R.~Chitra, Europhys. Lett. \textbf{63}, 440 
(2003). Note that in this work, some approximations were used 
valid only when the system is pinned for motion along the stripes, leading 
to results quite different than ours.

\bibitem{nelson} D. R. Nelson and J. M. Kosterlitz, Phys. Rev. Lett. \textbf{%
39}, 1201 (1977).

\bibitem{kubo} See article by R. Kubo, S.J. Miyake, and N. Hashitsume, in 
\textit{Solid State Physics} \textbf{17} (Academic Press, New York, 1965).

\bibitem{Fogler1} M. M. Fogler, in {\em High Magnetic Fields: Applications in 
Condensed Matter Physics and Spectroscopy}, ed. by C. Berthier, L.-P. Levy, 
G. Martinez (Springer-Verlag, Berlin, 2002).


\bibitem{exactsolution} V. L. Berezinskii, Sov. Phys. JETP \textbf{38}, 620
(1974).

\bibitem{footnotedisorderlevel} To use discrete fast-Fourier transformation
to compute the integrals over $\omega$ and $t$, we have to set a typical
stepsize in frequency to be much smaller than $e_x$ and $e_y$ and the
pinning peak widths. These quantities are dependent on the disorder level.
On the other hand, the frequency stepsize is inversely proportional to the
large-$t$ cutoff. As the depinning transition is approached and in the
depinned state that we will discuss in the next section, to see all the
interesting features at small $\omega$ we need to use very large cutoff. We
thus need to compromise on the disorder level in order to capture the
interesting physics at low frequencies.

\bibitem{dotsenko} V. Dotsenko, \textit{Introduction to the Replica Theory
of Disordered Statistical Systems,} (Cambridge, New York, 2001).


\bibitem{kohn} W. Kohn, Phys. Rev. \textbf{133}, A171 (1964).

\bibitem{chaikin} P.M. Chaikin and T.C. Lubensky, \emph{Principles of
Condensed Matter Physics} (Cambridge University Press, New York, 1995).

\bibitem{footnotedeltapeak} We want to mention that the weight of this delta
function term is not computed accurately in the current method since this
term has to be obtained along with the $\omega_n=0$ mode of $\tilde{\zeta}$.

\bibitem{missed} The anomalous finite frequency response for $\sigma
_{yy}(\omega )$ was not noticed in our early analysis of these results, and
so was not reported in Ref. \onlinecite{LFCY}.

\bibitem{sudbo} See, for example, S. Kragset, A. Sudbo, and F.S. Nogueira,
Phys. Rev. Lett. \textbf{92}, 186403 (2004).


\bibitem{Endo} A. Endo and Y. Iye, Phys. Rev. B 66, 075333 (2002).
\end{thebibliography}
\end{document}